\documentclass[12pt]{iopart}
\usepackage{iopams}
\usepackage{color} 
\usepackage{multicol}
\usepackage{graphicx}
\usepackage{placeins}
\usepackage{dcolumn}
\usepackage{bm}
\usepackage{subfigure}  
\usepackage{overpic}
\usepackage{slashbox}
\usepackage{ulem}
\usepackage{amsthm}
\usepackage{iopams}

\expandafter\let\csname equation*\endcsname\relax
\expandafter\let\csname endequation*\endcsname\relax
\usepackage{amsmath}
\usepackage{amssymb}
\usepackage{amsfonts}
\usepackage{tensor}
\usepackage{multirow}
\usepackage{makecell}
\usepackage{cite}
\newtheorem{thm}{Theorem}
\newcommand{\matr}[1]{\mathord{\buildrel{\lower3pt\hbox{$\scriptscriptstyle\leftrightarrow$}}\over{ \mathbf{#1}}}}
\begin{document}

\title{The Existence of Topological Edge States in Honeycomb Plasmonic Lattices}

\author{Li Wang,$^{1,\dagger}$ Ruo-Yang Zhang,$^{1,2,\dagger}$ Meng Xiao,$^{1,3}$ Dezhuan Han,$^4$ C. T. Chan$^1$ and Weijia Wen$^{1,*}$}

\address{$^1$Department of Physics, Hong Kong University of Science and Technology, Clear Water Bay, Hong Kong, China\\
$^2$Theoretical Physics Division, Chern Institute of Mathematics, Nankai University, Tianjin 300071, China\\
$^3$Department of Electrical Engineering, and Ginzton Laboratory, Stanford University, Stanford, California 94305, USA\\
$^4$Department of Applied Physics, Chongqing University, Chongqing 400044, China\\
$^\dagger$These authors contributed equally to this work}

\ead{$^*$phwen@ust.hk} 

\begin{abstract}
In this paper, we investigate the band properties of 2D honeycomb plasmonic lattices consisting of metallic nanoparticles. By means of the coupled dipole method and quasi-static approximation, we theoretically analyze the band structures stemming from near-field interaction of localized surface plasmon polaritons for both the infinite lattice and ribbons. Naturally, the interaction of point dipoles decouples into independent out-of-plane and in-plane polarizations. For the out-of-plane modes, both the bulk spectrum and the range of the momentum $k_{\parallel}$ where edge states exist in ribbons are similar to the electronic bands in graphene. Nevertheless, the in-plane polarized modes show significant differences, which do not only possess additional non-flat edge states in ribbons, but also have different distributions of the flat edge states in reciprocal space. For in-plane polarized modes, we derived the bulk-edge correspondence, namely, the relation between the number of flat edge states at a fixed $k_\parallel$, Zak phases of the bulk bands and the winding number associated with the bulk hamiltonian, and verified it through four typical ribbon boundaries, i.e. zigzag, bearded zigzag, armchair, and bearded armchair. Our approach gives a new topological understanding of edge states in such plasmonic systems, and may also apply to other 2D ``vector wave" systems.
\end{abstract}

%
\vspace{2pc}
\maketitle
%
%

\section{Introduction}
Recent progress in topological phases \cite{haldane1988model, zak1989berry, hatsugai1993chern, resta2000manifestations, hasan2010colloquium, qi2011topological, xiao2010berry} of condensed matters stimulates the studies of non-trivial topological properties in various physical systems \cite{lu2014topological, stanescu2010topological, xiao2015geometric, lubensky2015phonons}. In 2005, Haldane and Raghu introduced the topological concept in relation to the quantum Hall effect into photonic systems \cite{raghu2008analogs, haldane2008possible}. And then many physical phenomena in condensed matter physics which are difficult to be observed in the past have found their classical analogies \cite{wang2008reflection, wang2009observation, hafezi2013imaging, khanikaev2013photonic, chen2014experimental, bliokh2015quantum, lu2015experimental}. Artificial graphene \cite{polini2013artificial} is one of the typical examples, which offers a platform to investigate interesting phenomena related to the massless Dirac fermion dynamics near the Dirac cone\cite{pal2011dirac}. As shown by recent studies, various physical systems possess the spectra similar to the $\pi$ bands of electrons in graphene \cite{neto2009electronic, wakabayashi2010electronic}, such as 2D electron gas in quantum well \cite{gibertini2009engineering}, ultra cold atoms in optical lattices \cite{wu2008p}, photonic crystals\cite{ochiai2010topological, huang2011dirac, yannopapas2011gapless, bellec2013tight, plotnik2014observation, rechtsman2013topological, bellec2014manipulation}, and plasmonic lattices composed of metal particles \cite{han2009dirac, weick2013dirac}. Although $\pi$ bands in graphene are merely composed of $p_z$ orbital with only one freedom, it is also possible to construct band structures with multiple degrees of freedom and go beyond the ``scalar wave" nature in some systems with honeycomb lattice. A simple idea to achieve this goal is to build ``vector wave" bands by including the $p_{x,y}$ orbital of the lattice. Such schemes have been discussed for ultra cold atoms \cite{wu2008p}, polaritons of coupled micropillars \cite{jacqmin2014direct}, the TE (or in-plane polarized) modes in photonic systems\cite{han2009dirac, ochiai2010topological}, and the in-plane vibration modes in classical spring-mass systems \cite{wang2015coriolis}.

More interestingly, there exist symmetry-protected edge states between the upper and lower $\pi$ bands in graphene-like ribbon systems \cite{nakada1996edge, brey2006electronic, akhmerov2008boundary}, which have been confirmed experimentally both in graphene \cite{zhang2005experimental} and artificial graphene systems\cite{kuhl2010dirac, plotnik2014observation, bellec2014manipulation}. According to previous theoretical studies \cite{ryu2002topological, delplace2011zak, mong2011edge, hatsugai2009bulk}, the emergence of edge states is associated with the chiral symmetry of the bulk Hamiltonian and can be determined by the Zak phase of the $\pi$ bands, which is well-known as bulk-edge correspondence \cite{xiao2010berry, xiao2014surface, huang2014sufficient, graf2013bulk, mong2011edge, hatsugai2009bulk}. However, previous analyses of the ``vector wave" (in-plane polarized) bands of honeycomb lattices showed neither the existence ranges nor the number of the midgap edge states is identical with those of graphene \cite{han2009dirac, ochiai2010topological}, although they both possess the Dirac-like dispersion at the vertices of the first Brillouin zone. This is a strong hint that scalar wave bands and vector wave bands possess different topological properties, and the principle of bulk-edge correspondence for ``scalar waves" should be modified in order to be used for the multifreedom systems. Therefore, a proper relationship of edge states and topological properties of the bulk bands of ``vector wave" system is really needed.

In this paper, we mainly investigate the in-plane polarized modes in honeycomb plasmonic lattices (HPLs), as a representative example of vector wave systems. The interaction is simplified to form a concise Hermitian eigenvalue problem by using the coupled dipole method (CDM) and neglecting the dissipation and the retardation of the fields. Band structures of the infinite lattice and ribbons with four typical boundaries for out-of-plane and in-plane polarizations are calculated respectively in the tight-binding approximation. And the results taken the retardation and long range interactions into consideration also verify the effectiveness of our simplified model. It is shown that the ribbons with different boundary conditions possess different distributions of the edge states. Moreover, we develop a theorem of bulk-edge correspondence for the in-plane polarized modes which reveals the relations between the Zak phase, the winding number associated with bulk Hamiltonian and the existence of the zero-eigenvalue flat edge states in ribbons with various boundaries. Our result could be regarded as an extension of the theorem for graphene\cite{delplace2011zak}. The existences of edge states predicted by our theorem are perfectly consistent with the calculated band structures in the four typical ribbons respectively.

\section{\label{sec1:level1}Honeycomb plasmonic lattice and different boundaries\protect}
\begin{figure*}[t]
\centering
\includegraphics[width=1.0\linewidth,height=0.288\linewidth]{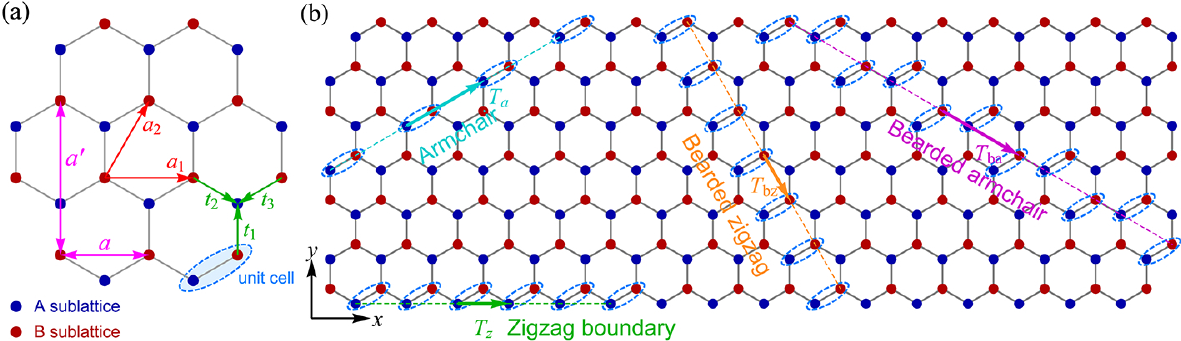}
\caption{\label{fig:1} (a) Schematic diagram of honeycomb lattice. Blue and red dots denote the two inequivalent sublattices A and B respectively and the light blue ellipse circles the unit cell (dimer) of lattice. $\bm{a}_1$, $\bm{a}_2$ are two basis vectors of the Bravais lattice. $\bm{t}_1$, $\bm{t}_2$, $\bm{t}_3$ are the three vectors connecting the nearest neighbours. $a=|\bm{a}_1|=|\bm{a}_2|= \sqrt 3 {a_0}$ and $a'=\frac{3}{2}{a_0}$ are the periods along $x$ and $y$ directions respectively. (b) Four different edge types, i.e. zigzag, bearded zigzag, armchair, bearded armchair, obtained by cutting lattice in different directions. }
\end{figure*}
Metal nanoparticle lattice is a particular type of photonic systems which supports the collective resonance of localized surface plasmons of individual particles and have distinctive band dispersions \cite{wang2001three, maier2003optical, weber2004propagation, park2004surface, fung2007plasmonic, zhen2008collective, fung2008analytical}. Several studies have reported the interesting topological phenomena emerging in this type of systems \cite{han2009dirac, poddubny2014topological, sinev2015mapping, ling2015topological, ling2015formation}. Specially, 2D honeycomb plasmonic lattices are made up of one-layer-thick metal particles and arranged in honeycomb structure just like graphene, which can be regarded as a compound structure nested with two triangular sublattices. The Bravais lattice is shown in Fig.~\ref{fig:1}(a), where $\bm{a}_1$, $\bm{a}_2$ are two basis vectors, and two neighboring ``atoms" consist of a unit cell (dimer). And a ribbon is a lattice with finite width in one direction with two parallel edges and is infinite along the direction of edges. Actually, there are myriad different types of ribbon edges, and a large class of ribbon edges can be characterized by the periodic vector $\mathbf{T}(m,n)=m\bm{a}_1+n\bm{a}_2$, where $(m,n)$ is a pair of coprime integers \cite{delplace2011zak}. According to Delplace's work \cite{delplace2011zak}, to obtain the boundary of a ribbon with the periodic vector $\mathbf{T}(m,n)$, we should firstly translate a dimer with the displacement $\bm{a}_1$ ($-\bm{a}_1$ if $m<0$) $|m|$ times (each time, we obtain a dimer on the edge), then translate with $\bm{a}_2$ ($-\bm{a}_2$ if $n<0$) $|n|$ times, and so on and so forth. There are three different orientations for a unit cell, the one adopted in our paper is shown in Fig.~\ref{fig:1}(a), the other two can be obtained from the former by a $\pm 2\pi /3$ rotation. For infinite systems, there are no differences between these three choices except an additional phase factor of the bulk Hamiltonian. However, they give rise to different results in finite systems even constructing the edge of a ribbon with the same $\mathbf{T}(m,n)$, in addition, the results are also different when calculating the Zak phases and winding numbers of bulk bands. Specifically, Fig.~\ref{fig:1}(b) shows the most studied four types of ribbon boundaries, i.e. zigzag, armchair, bearded zigzag, and bearded armchair. One choice (but not unique) of the periodic vectors of these four boundaries in Fig.~\ref{fig:1}(b) are, respectively,
\begin{align*}
\text{Zigzag:} &\qquad \mathbf{T}_\mathrm{z} \ =\mathbf{T}(1,0),\\
\text{Armchair:} &\qquad \mathbf{T}_\mathrm{a} \ =\mathbf{T}(1,1),\\
\text{Bearded zigzag:} &\qquad \mathbf{T}_\mathrm{bz}=\mathbf{T}(1,-1),\\
\text{Bearded armchair:} &\qquad \mathbf{T}_\mathrm{ba}=\mathbf{T}(2,-1).\\
\end{align*}

\section{\label{sec2:level1}Band structure and edge states\protect}

\subsection{Quasi-static and dipole approximation}
In this paper, we use CDM to simplify the interactions between metal particles with treating metal particles as point dipoles, which means the results are effective only for ${r_s}/{a_0} \le 1/3$ \cite{maier2003optical, weber2004propagation, park2004surface}, where $r_s$ and $a_0$ are the particle radius and the center-to-center distance between two adjacent particles respectively. The electric field generated by a harmonically radiating point dipole is characterized by the equation
\begin{equation}
\mathbf{E}(\mathbf{r}) = \frac{1}{{4\pi {\varepsilon _0}}}\left\{ {k^2}(\mathbf{n} \times \mathbf{P}) \times \mathbf{n}\frac{{{e^{ikr}}}}{r} \right.
\left.+ [3\mathbf{n}(\mathbf{n} \cdot \mathbf{P}) - \mathbf{P}] \left( \frac{1}{{{r^3}}} - \frac{{ik}}{{{r^2}}} \right) {e^{ikr}}\right\} ,
\label{eq:1}
\end{equation}
where $\mathbf{r} $ is the displacement vector from the source point to the field point, $\mathbf{n}$ is the unit vector along $\mathbf{r}$, $\mathbf{P}$ is the dipole moment,  $k=\omega/c$ is the wave vector (here $\omega$ is the angular frequency), and the surrounding medium is assumed to be vacuum. When the wavelength $\lambda  \gg {a_0}$, the quasi-static approximation (QSA) can be applied, then Eq.~(\ref{eq:1}) reduces to
\begin{equation}
\mathbf{E}(\mathbf{r}) = \frac{1}{{4\pi {\varepsilon _0}}}\frac{{3\mathbf{n}(\mathbf{n} \cdot \mathbf{P}) - \mathbf{P}}}{{{r^3}}} = \frac{1}{{4\pi {\varepsilon _0}}}\mathbf{\mathord{\buildrel{\lower3pt\hbox{$\scriptscriptstyle\leftrightarrow$}}\over G}} (\mathbf{r}) \cdot \mathbf{P},
\label{eq:2}
\end{equation}
with the quasi-static Green tensor
\begin{equation}
\mathbf{\mathord{\buildrel{\lower3pt\hbox{$\scriptscriptstyle\leftrightarrow$}}
\over G}} (\mathbf{r}) = \frac{{3\mathbf{n} \otimes \mathbf{n} - \mathbf{\mathord{\buildrel{\lower3pt\hbox{$\scriptscriptstyle\leftrightarrow$}}
\over I}} }}{{{r^3}}},
\label{eq:3}
\end{equation}
where $\mathbf{\mathord{\buildrel{\lower3pt\hbox{$\scriptscriptstyle\leftrightarrow$}} \over I}}$ is identity matrix. For simplicity, we assume the metal particles are all spherical. The induced dipole moment of the particle on the lattice point $\mathbf{R}$ (Coupled Dipole Equation) is
\begin{equation}
\begin{array}{l}
{\mathbf{P}_\mathbf{R}} = \alpha (\omega )\sum\limits_{\mathbf{R}' \ne \mathbf{R}} {{\mathbf{E}_{\mathbf{R}',\mathbf{R}}}}  = \tilde \alpha (\omega )\sum\limits_{\mathbf{R}' \ne \mathbf{R}} {\mathbf{\mathord{\buildrel{\lower3pt\hbox{$\scriptscriptstyle\leftrightarrow$}}
\over G}} (\mathbf{R} - \mathbf{R}') \cdot {\mathbf{P}_{\mathbf{R}'}}} = \tilde \alpha (\omega )\sum\limits_{\mathbf{R}' \ne \mathbf{R}} {{{\mathbf{\mathord{\buildrel{\lower3pt\hbox{$\scriptscriptstyle\leftrightarrow$}}
\over G}} }_{\mathbf{R},\mathbf{R}'}} \cdot {\mathbf{P}_{\mathbf{R}'}}},
\end{array}
\label{eq:4}
\end{equation}
where $\mathbf{E}_{\mathbf{R}',\mathbf{R}}$ denotes the electric field at $\mathbf{R}$ generated by the dipole on the lattice point $\mathbf{R}'$, $\tilde \alpha (\omega ) = \alpha (\omega )/(4\pi {\varepsilon _0)} = \frac{{\varepsilon (\omega ) - 1}}{{\varepsilon (\omega ) + 2}}r_s^3$ is the polarizability of the spherical particles, ${\varepsilon (\omega )}$ is the frequency-dependent relative permittivity of the metal particles, and $\mathbf{{\mathord{\buildrel{\lower3pt\hbox{$\scriptscriptstyle\leftrightarrow$}}
\over G}} _{\mathbf{R},\mathbf{R}'}} = \mathbf{\mathord{\buildrel{\lower3pt\hbox{$\scriptscriptstyle\leftrightarrow$}}
\over G}} (\mathbf{R} - \mathbf{R}')$ is the two-point correlation function. Indeed, the coupled dipole equation can be transformed to an eigenvalue problem
\begin{equation}
\sum\limits_{\mathbf{R}' \ne \mathbf{R}} {{{\mathbf{\mathord{\buildrel{\lower3pt\hbox{$\scriptscriptstyle\leftrightarrow$}}
\over G}} }_{\mathbf{R},\mathbf{R}'}} \cdot {\mathbf{P}_{\mathbf{R}'}}}  = \tilde \alpha {(\omega )^{ - 1}}{\mathbf{P}_{\mathbf{R}'}},
\label{eq:5}
\end{equation}
where $\tilde{\alpha}(\omega)^{-1}$ serves as the eigenvalue.
\subsection{Eigenvalue problem and Hamiltonian}
Since the honeycomb lattice is a compound lattice with two inequivalent sublattices (see Fig.~\ref{fig:1}), we can combine the dipole moments of two sublattice points in one unit cell to a single polarization vector ${\mathbf{P}_\mathbf{R}} = {(\mathbf{P}_\mathbf{R}^A,\mathbf{P}_\mathbf{R}^B)^T}$, then Eq.~(\ref{eq:5}) can be written as an Hermitian eigenvalue equation
\begin{equation}
\sum\limits_{\mathbf{R}'} {{{H }_{\mathbf{R},\mathbf{R}'}} \cdot {\mathbf{P}_{\mathbf{R}'}}}  = \tilde \alpha {(\omega )^{ - 1}}{\mathbf{P}_{\mathbf{R}'}},
\label{eq:6}
\end{equation}
where $\mathbf{R}(\mathbf{R}')$ denotes the index of a unit cell, the Hamiltonian element $H_{\mathbf{R},\mathbf{R}'} = H(\mathbf{R} - \mathbf{R}')$ correlating the two cells $\mathbf{R}$ and $\mathbf{R}'$ has the block form
\begin{equation}
H(\Delta \mathbf{R}) = \left( {\begin{array}{*{20}{c}}
{{H_{AA}}(\Delta \mathbf{R})}&{{H_{AB}}(\Delta \mathbf{R})}\\
{{H_{BA}}(\Delta \mathbf{R})}&{{H_{BB}}(\Delta \mathbf{R})}
\end{array}} \right),
\label{eq:7}
\end{equation}
with $\Delta \mathbf{R} = \mathbf{R} - \mathbf{R}'$. The hopping terms $H_{AA}$ and $H_{BB}$ represent the interaction from the same sublattices $A$ and $B$ respectively, between different unit cells, and they take the forms $H_{AA}(0)=H_{BB}(0)=\mathbf{0}$, $H_{AA}(\Delta \mathbf{R})=H_{BB}(\Delta \mathbf{R})=\mathbf{\mathord{\buildrel{\lower3pt\hbox{$\scriptscriptstyle\leftrightarrow$}}
\over G}} (\Delta \mathbf{R})(\Delta \mathbf{R} \ne 0)$. Similarly, the hopping terms $H_{AB}$ and $H_{BA}$ represent respectively the interaction of $B$ to $A$ and $A$ to $B$, and they read ${H_{AB}}(\Delta \mathbf{R}) = {H_{BA}}( - \Delta \mathbf{R}) = \mathbf{\mathord{\buildrel{\lower3pt\hbox{$\scriptscriptstyle\leftrightarrow$}}
\over G}} (\Delta \mathbf{R} - {\mathbf{t}_3})$. Further, by introducing the creation and annihilation operators, we can write the Hamiltonian in a second quantized form
\begin{equation}
H = \sum\limits_{\mathbf{R},\mathbf{R}'} {(\bm{a}_\mathbf{R}^\dag ,\bm{b}_\mathbf{R}^\dag )} {H_{\mathbf{R},\mathbf{R}'}}{({\bm{a}_{\mathbf{R}'}},{\bm{b}_{\mathbf{R}'}})^T}.
\label{eq:8}
\end{equation}
Eq.~(\ref{eq:6}) and Eq.~(\ref{eq:7}) give the eigenequation and the component form of the Hamiltonian in Wannier basis $\left| \mathbf{R} \right\rangle $. To consider the common eigenstates of both the Hamiltonian and the lattice translation operator subjected to the periodic boundary condition, we can obtain the Bloch state $|\psi_{\mathbf{k}}\rangle=\mathbf{P}_{\mathbf{R}}(\mathbf{k})|\mathbf{R}\rangle$ with the component $\mathbf{P}_{\mathbf{R}}(\mathbf{k})=e^{i\mathbf{k}\cdot\mathbf{R}}\psi_{\mathbf{k}}$, where $\mathbf{k}$ is the Bloch wave vector corresponding to the eigenvalue of the lattice translation operator. On the basis of Bloch states, the eigenequation of the Hamiltonian becomes $ H_{\mathbf{k}}\cdot\psi_{\mathbf{k}}=\tilde\alpha(\omega)^{-1}\psi_{\mathbf{k}}$ with the component of the Hamiltonian in Bloch representation $ H_{\mathbf{k}}=\sum_{\Delta\mathbf{R}}H(\Delta\mathbf{R})e^{-i\mathbf{k}\cdot\Delta\mathbf{R}}$. According to the tight-binding model (TBM), the Hamiltonian with only the nearest-neighbour interaction is
\begin{equation}\label{eq:9}
  H_{\mathbf{k}}=\left(\begin{array}{c|c}
                       \mathbf{0} & \mathcal{A}(\mathbf{k})\\\hline
                        \mathcal{A}(\mathbf{k})^\dagger & \mathbf{0}
                       \end{array} \right),
\end{equation}
where $\mathcal{A}(\mathbf{k})=\sum_{i=1}^3\mathbf{\mathord{\buildrel{\lower3pt\hbox{$\scriptscriptstyle\leftrightarrow$}}\over G} } (\mathbf{t}_i)\,e^{-i\mathbf{k}\cdot(\mathbf{t}_i-\mathbf{t}_3)}$, $\mathbf{t}_i(i=1,2,3)$ denotes three nearest hopping vectors respectively (see Fig.~\ref{fig:1}). It is easy to know that the Hamiltonian possesses chiral symmetry: $C\cdot H_{\mathbf{k}}\cdot C=-H_\mathbf{k}$ where $C=\mathrm{diag}(\mathbf{\mathord{\buildrel{\lower3pt\hbox{$\scriptscriptstyle\leftrightarrow$}} \over I}}, -\mathbf{\mathord{\buildrel{\lower3pt\hbox{$\scriptscriptstyle\leftrightarrow$}} \over I}})$. This property indicates that there may exist zero-eigenvalue edge states in corresponding ribbon systems~\cite{ryu2002topological}.
Actually, the tight-binding Hamiltonian given in Eq.~(\ref{eq:9}) is quite similar to the $\pi$ electron Hamiltonian in graphene except the nearest neighbor hopping constant is replaced by the Green tensors $\mathbf{\mathord{\buildrel{\lower3pt\hbox{$\scriptscriptstyle\leftrightarrow$}}
\over G}}(\mathbf{t}_i)$. Nevertheless, this difference has remarkable influences on band structures and the topological properties of the edge states.

Being the eigenvalue of an Hermitian matrix, $\tilde\alpha(\omega)^{-1}$ is required to be real. However, if the permittivity $\varepsilon(\omega)$ of metal particles has imaginary part, a real $\tilde\alpha(\omega)$ would lead to a complex frequency $\omega$ which means that the plasmonic resonance of the particles would decay over time. Without changing the topological properties, the imaginary part of the permittivity is omitted and the Drude model without damping $\varepsilon(\omega)=1-\omega_p^2/\omega^2$ is used to characterize the metal paricles, where $\omega_p$ is the plasmonic resonant frequency. Then we can obtain the relation between $\tilde\alpha$ and $\omega$
\begin{equation}\label{eq:10}
  \tilde{\alpha}(\omega)=\frac{\omega_0^2}{\omega_0^2-\omega^2}{r_s}^3,
\end{equation}
where $\omega_0=\omega_p/\sqrt{3}$ is the dipole resonant frequency of spheres. For $\omega>0$, $\tilde{\alpha}(\omega)$ is a monotonic function, thus the correspondence between the eigenvalues and eigenfrequencies in Eq.~(\ref{eq:6}) is a one-to-one mapping\cite{ling2015topological}.

\subsection{Bulk band structures of HPLs}

Under the previous simplification, the band structure of honeycomb plasmonic lattice can be easily obtained from Eq.~(\ref{eq:6}). Because $H_{\mathbf{k}}$ is a $6 \times 6$ matrix, the system has $6$ bands in total. On the other hand, the in-plane polarization $\mathbf{P}_{xy}$ of the dipoles is orthogonal and decoupled with the out-of-plane polarization $P_z$, since the Green tensor can be decomposed as the direct sum of the in-plane and out-of-plane parts $\mathbf{\mathord{\buildrel{\lower3pt\hbox{$\scriptscriptstyle\leftrightarrow$}}
\over G}} (\mathbf{r}) = {\mathbf{\mathord{\buildrel{\lower3pt\hbox{$\scriptscriptstyle\leftrightarrow$}}
\over G}} _{xy}}(\mathbf{r}) \oplus {G_z}(\mathbf{r})$ with ${{\mathbf{\mathord{\buildrel{\lower3pt\hbox{$\scriptscriptstyle\leftrightarrow$}}
\over G}} }_{xy}}(\mathbf{r}) = (3\mathbf{n} \otimes \mathbf{n} - {{\mathbf{\mathord{\buildrel{\lower3pt\hbox{$\scriptscriptstyle\leftrightarrow$}}
\over I}} }_{xy}})/{r^3}$ and ${G_z}(\mathbf{r}) =  - 1/{r^3}$ for all $\mathbf{r}$ in $xy$ plane. Therefore, it makes physics more clear if we investigate the out-of-plane and in-plane polarized eigenspectra of the two directly summed parts of the Hamiltonian $H_{\mathbf{k}}=H^{\mathrm{in}}_{\mathbf{k}}\oplus H^{\mathrm{out}}_{\mathbf{k}}$ separately. The out-of-plane part of the Hamiltonian has a 2D massless Dirac-like form:
\begin{equation}\label{eq:11}
H^\mathrm{out}_\mathbf{k}=\left(\begin{array}{cc} 0 & p(\mathbf{k}) \\ p(\mathbf{k})^* & 0\end{array}\right)=\mathbf{p}(\mathbf{k})\cdot \bm{\sigma},
\end{equation}
where $p(\mathbf{k})=-\frac{1}{a_0^3}\sum_{i=1}^3 e^{-i\mathbf{k}\cdot(\mathbf{t}_i-\mathbf{t}_3)}$, $\bm{\sigma}=(\sigma_x,\sigma_y,\sigma_z)$ is the triplet of Pauli matrices, and here $ \mathbf{p}(\mathbf{k})
=(\mathrm{Re} \ p(\mathbf{k}),-\mathrm{Im} \ p(\mathbf{k}), 0)$. It has two eigenvalues
\begin{equation}\label{eq:12}
\alpha_{\pm}(\omega)^{-1}(\mathbf{k})=\pm|p(\mathbf{k})|,
\end{equation}
and the corresponding eigenstates are
\begin{equation}\label{eq:13}
\psi_\pm(\mathbf{k})=\frac{1}{\sqrt{2}}
\begin{pmatrix}
e^{i\mathrm{Arg}[p(\mathbf{k})]} \\ \pm 1
\end{pmatrix}.
\end{equation}
The Hamiltonian of the in-plane modes is a $4$ dimensional matrix
\begin{equation}
H^{\mathrm{in}}_{\mathbf{k}} = \left( {\begin{array}{*{20}{c}}
0&A(\mathbf{k})\\
{{A^\dag(\mathbf{k}) }}&0
\end{array}} \right),
\label{eq:14}
\end{equation}
where $A(\mathbf{k})=\sum_{i=1}^3\mathbf{\mathord{\buildrel{\lower3pt\hbox{$\scriptscriptstyle\leftrightarrow$}}
\over G}}_{xy}(\mathbf{t}_i)\,e^{-i\mathbf{k}\cdot(\mathbf{t}_i-\mathbf{t}_3)}$ is a $2$ dimensional symmetric matrix. The solutions of the eigenequation $H^{\mathrm{in}}_{\mathbf{k}}\psi=\lambda\psi$ have the form
\begin{equation}
\psi  = \frac{1}{{\sqrt 2 }}\left( \begin{array}{c}
{e^{ - i\gamma }}\mathbf{u}\\
{\mathbf{u}^*}
\end{array} \right),
\label{eq:15}
\end{equation}
where $\psi$ and $\mathbf{u} $ are both normalized, and $e^{-i\gamma}$ is an additional phase, moreover, $\mathbf{u}$ satisfies the corresponding con-eigenvalue problem
\begin{equation}
A{\mathbf{u}^*} = \lambda {e^{ - i\gamma }}\mathbf{u} = {\tilde \lambda ^*}\mathbf{u},
\label{eq:16}
\end{equation}
with $\tilde \lambda  = \lambda {e^{i\gamma }} = {\mathbf{u}^T}{A^\dag }\mathbf{u}$. In addition, $\psi$ is also the eigenvector of ${(H_{\mathbf{k}}^{\mathrm{in}})^2} = \mathrm{diag}(A{A^\dag },{A^\dag }A)$, thus $\mathbf{u}$ is the solution of the following equation
\begin{equation}
{H_{\mathrm{eff}}}\mathbf{u} = {\lambda ^2}\mathbf{u},
\label{eq:17}
\end{equation}
where the effective Hamiltonian ${H_{\mathrm{eff}}} = A{A^\dag }$ which can also be expressed in the Dirac-like form
\begin{equation}
{H_{\mathrm{eff}}} = A{A^\dag } = \sum\limits_{\alpha  = 0}^3 {{g^\alpha }{\sigma _\alpha }}, \quad ({g^\alpha } \in \mathbb{C}).
\label{eq:18}
\end{equation}
Naturally, we get two eigenvalues of $H_{\mathrm{eff}}$
\begin{equation}
\lambda _j^2 = {g^0} - {( - 1)^j}\left| \mathbf{g} \right|, \quad (j=1,2),
\label{eq:19}
\end{equation}
and the corresponding eigenvectors
\begin{equation}
\begin{array}{l}
{\mathbf{u}_1} = \frac{1}{{\sqrt 2 }}\left( \begin{array}{c}
\sqrt {1 + \cos \theta } \\
\sqrt {1 - \cos \theta } {e^{i\varphi }}
\end{array} \right), \quad {\mathbf{u}_2} = \frac{1}{{\sqrt 2 }}\left( \begin{array}{c}
\sqrt {1 - \cos \theta } {e^{ - i\varphi }}\\
 - \sqrt {1 + \cos \theta }
\end{array} \right),
\end{array}
\label{eq:20}
\end{equation}
where $\theta$, $\varphi$ are the components of $\mathbf{g} = (\left| \mathbf{g} \right|,\theta ,\varphi )$ in spherical coordinates. Note that $\varphi$ is not well-defined when $\mathbf{g}$ is along $z$ axis, hence the phase $e^{i\varphi}$($e^{-i\varphi}$) in Eq.~(\ref{eq:20}) is chosen to ensure the continuity of $\mathbf{u}_1$ ($\mathbf{u}_2$) at $g_x=g_y=0$. According to Eq.~(\ref{eq:19}), there are four eigenvalues of the ordinary Hamiltonian Eq.~(\ref{eq:14}):
\begin{equation}
{\tilde \alpha _{j, \pm }}{(\mathbf{k})^{ - 1}} = {\lambda _{j, \pm }}(\mathbf{k}) =  \pm \sqrt {{g^0}(\mathbf{k}) - {{( - 1)}^j}\left| {\mathbf{g}(\mathbf{k})} \right|},
\label{eq:21}
\end{equation}
which represent the four bulk bands of the in-plane modes. And the eigenvectors of these four bands are, respectively,
\begin{equation}
{\psi _{j, \pm }}(\mathbf{k}) = \frac{1}{{\sqrt 2 }}\left( \begin{array}{c}
 \pm {e^{ - i{\gamma _j}(\mathbf{k})}}{\mathbf{u}_{j}}(\mathbf{k})\\
{\mathbf{u}_{j}}{(\mathbf{k})^*}
\end{array} \right), \quad (j=1,2),
\label{eq:22}
\end{equation}
where $\gamma_j$ can be determined by the argument of ${\tilde \lambda _j} \ ({\tilde \lambda _j} = \mathbf{u}_j^T{A^\dag }{\mathbf{u}_j})$.
\begin{figure}[tb]
\centering
\includegraphics[width=0.8\linewidth,height=0.4285\linewidth]{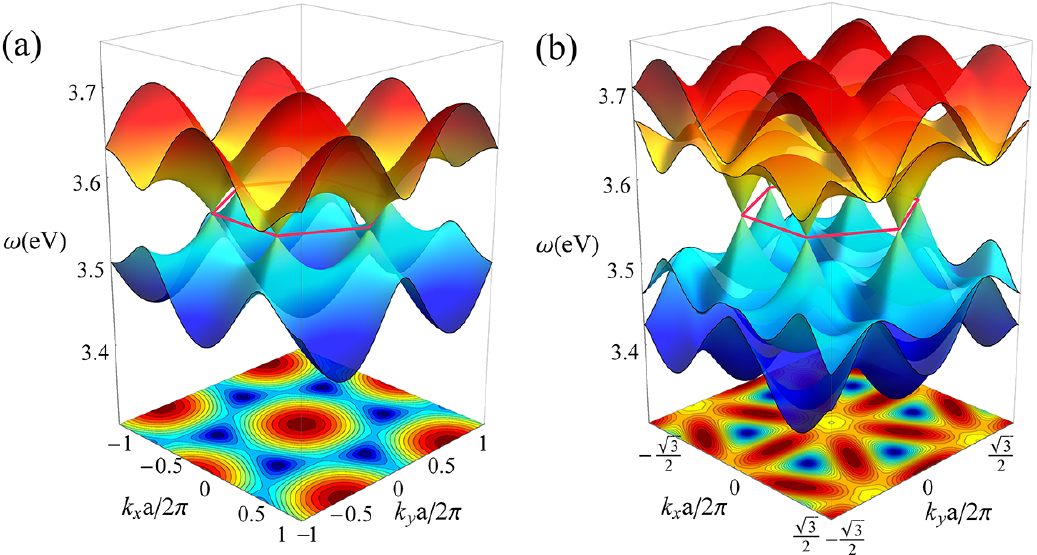}
\caption{\label{fig:2} Bulk band structures, (a) out-of-plane polarized modes, (b) in-plane polarized modes with QSA and TBM, where the energy is in unit of eV, the band number is labeled as 1, 2 (1, 2, 3, 4) from bottom to top for (a) (and (b)), undersurface patterns are the projections of band 2 in (a) and band 3 in (b) respectively.}
\end{figure}
The bulk band structures of out-of-plane and in-plane polarizations are shown in Figs.~\ref{fig:2}(a) and ~\ref{fig:2}(b) respectively. Here and in the following calculations, the parameters are chosen to be ${\omega _p} = 6.18$ eV, $r_s=10\ \mathrm{nm}$, and $r_s/a_0=\sqrt{3}/6$. Both of the in-plane and out-of-plane bands possess Dirac cones located at the $\mathbf{K}$ and $\mathbf{K}'$ points (the corners of first Brillouin zone), and all of the Dirac points have the same frequency $\omega_0$ corresponding to the zero-eigenvalue ($\tilde \alpha {({\omega _0})^{ - 1}} = 0$). For the out-of-plane case, the eigenvector on each ``atom" has a fixed polarization along $z$ direction just like the scalar wave function composed of $p_z$ orbital of electrons in graphene, therefore the out-of-plane polarized band of honeycomb plasmonic lattices pretty resembles the $\pi$ bands of electron in graphene and acts as a classical photonic analog. Meanwhile, the in-plane polarized modes have two polarization degrees of freedom, this vector nature of eigenfield on all lattice points leads to its four-band peculiarity as shown in Fig.~\ref{fig:2}(b). However, in graphene, the $p_{xy}$ orbital hybridizes with the $s$ orbital, composing the $\sigma$ band, and its anisotropic feature is not so conspicuous because of the large $s$-orbital component. Moreover, the $\sigma$ band is fully filled and inert. Therefore, the similar ``vector wave'' excitation is not observed in graphene. Another property of the in-plane bands is that the upper two bands and the lower two bands joint with a quadratic form at $\Gamma$ point. In addition, it is obvious that both the out-of-plane and the in-plane band structures are symmetric about $\omega_0$ plane due to chiral symmetry under the nearest neighbour tight-binding approximation.
\subsection{Band structures and edge states of ribbons}
In this section, we investigate the band structures of plasmonic ribbons with four types of boundary conditions, i.e. zigzag, bearded zigzag, armchair, and bearded armchair. For a finite ribbon with width $N$, the Hamiltonians are $2N$ order and $4N$ order square matrices for out-of-plane and in-plane polarizations respectively. With accuracy to the nearest neighbor hopping, we can obtain the Hamiltonians for zigzag and armchair ribbons respectively
\begin{align}
{H_{\text{zig}}} =& \sum\limits_{{k_\parallel }} {\left\{ {\sum\limits_{m \in \text{odd}} {a _{{k_\parallel }}^\dag (m)(G_3 + G_2{e^{ - i{k_\parallel }a}})} {b _{{k_\parallel }}}(m)} \right.} \notag \\&+ \sum\limits_{m \in \text{even}} {a _{{k_\parallel }}^\dag (m)(G_2 + G_3{e^{i{k_\parallel }a}})} {b _{{k_\parallel }}}(m)
\left. +{\sum\limits_{m = 2}^N {a _{{k_\parallel }}^\dag (m)G_1{b _{{k_\parallel }}}(m - 1) + {\rm{H}}{\rm{.c}}{\rm{.}}} } \right\},\label{eq:23}\\
{H_{\text{arm}}} =& \sum\limits_{{k_\parallel }} {\left\{ {\sum\limits_{m \in \text{odd}} {a _{{k_\parallel }}^\dag (m)G_1{e^{ - i{k_\parallel }a'}}} {b _{{k_\parallel }}}(m)} \right.}  + \sum\limits_{m \in \text{even}} {a _{{k_\parallel }}^\dag (m)G_1} {b _{{k_\parallel }}}(m) \notag \\
&+\sum\limits_{m = 1}^{N - 1} {a _{{k_\parallel }}^\dag (m)G_3{b _{{k_\parallel }}}(m + 1)} \left. +{\sum\limits_{m = 2}^N {a _{{k_\parallel }}^\dag (m)G_2{b _{{k_\parallel }}}(m - 1) + {\rm{H}}{\rm{.c}}{\rm{.}}} } \right\},\label{eq:24}
\end{align}
where $k_{\parallel}$ is the Bloch wave vector alone the periodic direction, and $G_i$ should be substituted by $-1/{a_0^3}$ for out-of-plane case, and by $\mathbf{\mathord{\buildrel{\lower3pt\hbox{$\scriptscriptstyle\leftrightarrow$}}
\over G}}_{xy}(\mathbf{t}_i)$ for in-plane case. And the Hamiltonians of bearded zigzag and bearded armchair ribbons are almost the same as their non-bearded counterparts except one more hopping term from the beard atoms of each edge. Immediately, we can obtain the band structures by solving the eigenvalue equations of these Hamiltonian matrices.
\begin{figure}[tb]
\centering
\includegraphics[width=0.8\linewidth,height=0.69\linewidth]{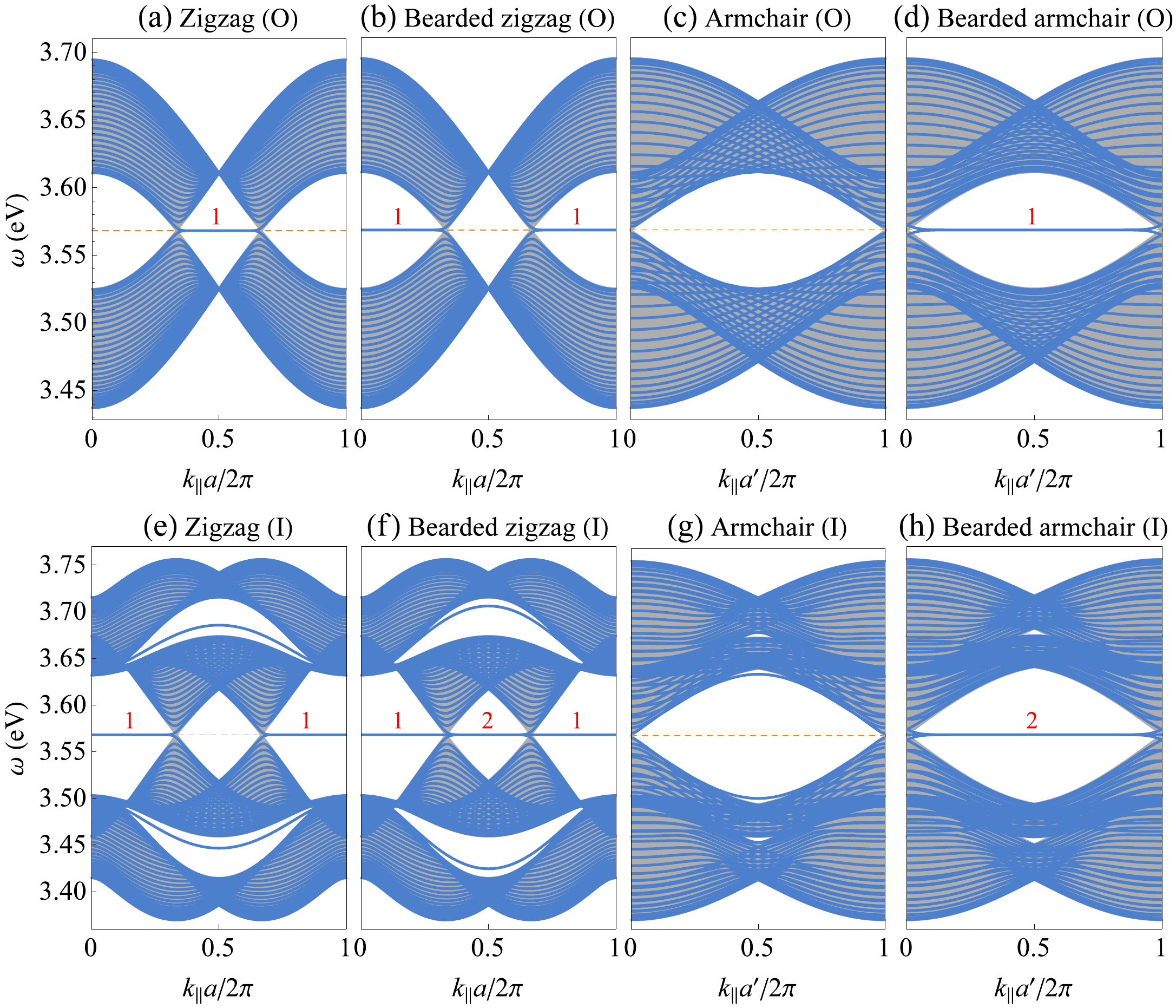}
\caption{\label{fig:3} Out-of-plane (O) and in-plane (I) polarized band structures (blue curves) of ribbons and projections of the bulk bands (gray region) with QSA and TBM. These $8$ subfigures correspond to $4$ different boundaries, i.e. (a)(e) zigzag ribbon, (b)(f) bearded zigzag ribbon, (c)(g) armchair ribbon, (d)(h) bearded armchair ribbon. The numbers of pairs of flat edge states are labeled above the corresponding flat edge bands. Orange dashed line is the contour of the dipole resonance frequency $\omega_0=3.568$ eV, and the ribbon width is chosen to be $N = 30$ for all of the eight cases.}
\end{figure}

Fig.~\ref{fig:3} shows the band spectra of out-of-plane and in-plane polarized modes with four different types of ribbon boundaries. The ribbon band structures have some generic properties for both out-of-plane and in-plane cases. To be more specific, most of the blue lines lie in the projected bulk band area (gray region) with some additional bands outside the projected band which attribute to edge states. The blue lines tend to fill up the whole gray region as the widths of the ribbons increase towards infinity. As the boundaries at two sides of a ribbon are kept to be the same, the edge bands always appear in pairs. In each pair, there are two nearly overlapped bands that are excited on different sides of the ribbon. In addition, there exists flat edge dispersion connecting the projected Dirac points at the mid frequency $\omega_0$. For the out-of-plane case, each spectrum of the four types is similar to the corresponding graphene ribbon. Zigzag ribbons possess a pair of edge bands in the interval $1/3 \sim 2/3$ of the first Brillouin zone, whereas bearded zigzag ribbons possess a pair of edge bands in the complementary region. Armchair ribbons do not show edge states, but bearded armchair ribbons possess edge states over the entire Brillouin zone. However, band structures of in-plane polarized modes reveal many remarkable differences from the out-of-plane case. Firstly, there exist not only flat edge states at $\omega_0$, but also two pairs of almost overlap bent edge bands in the gaps between bands $1,2$ and between bands $3,4$ for both zigzag and bearded zigzag ribbons. Besides, both the numbers and the locations in reciprocal space of flat edge states are different from those of the out-of-plane case with the same type of edge. For zigzag ribbons, a pair of edge bands appears within the complementary region, i.e. the interval $[0,1/3) \cup (2/3,1]$ of the Brillouin zone, in comparison with the out-of-plane case. For bearded zigzag ribbons, there exists not only a pair of flat edge bands in the interval $[0,1/3) \cup (2/3,1]$ but also two pairs in complementary region $[1/3,2/3)$. For bearded armchair ribbons, two pairs of flat edge bands pass through the whole Brillouin zone. The distributions of the flat edge states in all of these cases are summarized in Table I.
\begin{table}[htb]
\centering
\caption{Distribution of flat edge states. In the table, the interval denotes the range of $k_{\parallel}$ in the first Brillouin zone (with the normalized width) where the edge states appear.}
\begin{tabular}{l|cc|cc}
\hline
      &      \multicolumn{2}{c|}{Out-of-plane}  & \multicolumn{2}{c}{In-plane}    \\\cline{2-5}
  &\rule[-1pt]{0pt}{10pt}   pairs & interval  & pairs & interval \\ \hline
\multirow{2}*{Zigzag} &     \multirow{2}*{1} & \multirow{2}*{$(1/3,\, 2/3)$}  &    \multirow{2}*{1} & \multirow{2}*{$[0,\, 1/3)\cup(2/3,\, 1]$}\\  & & & &   \\ \hline
\multirow{2}*{\makecell[l]{Bearded zigzag}}   &    \multirow{2}*{1} & \multirow{2}*{$[0,\, 1/3)\cup(2/3,\, 1]$}  \rule[0pt]{0pt}{10pt}&  1 & $[0,\, 1/3)\cup(2/3,\, 1]$    \\
                                                     &    \rule[-5pt]{0pt}{10pt} &   &  2 & $(1/3,\, 2/3)$    \\ \hline
\multirow{2}*{Armchair} &    \multirow{2}*{0} &       & \multirow{2}*{0} & \\ & & & & \\ \hline
\multirow{2}*{\makecell[l]{Bearded armchair}}   &   \multirow{2}*{1} & \multirow{2}*{$(0,\, 1)$}  & \multirow{2}*{2} & \multirow{2}*{$(0,\, 1)$}\\
 & & & & \\
\hline
\end{tabular}\label{table1}
\end{table}
By checking the eigenvectors, the additional flat bands in the whole Brillouin zone for bearded zigzag and bearded armchair arise from the localized resonances at the beards. Another notable effect is that both the out-of-plane and in-plane modes of armchair ribbons are metallic, namely the gap at $\omega_0$ is strictly closed at the Dirac points, only when ribbon width $N=3M-1$ ($M \in \mathbb{N}$), and this is same to the electronic graphene system.

\begin{figure}[tb]
\centering
\includegraphics[width=0.8\linewidth,height=0.664\linewidth]{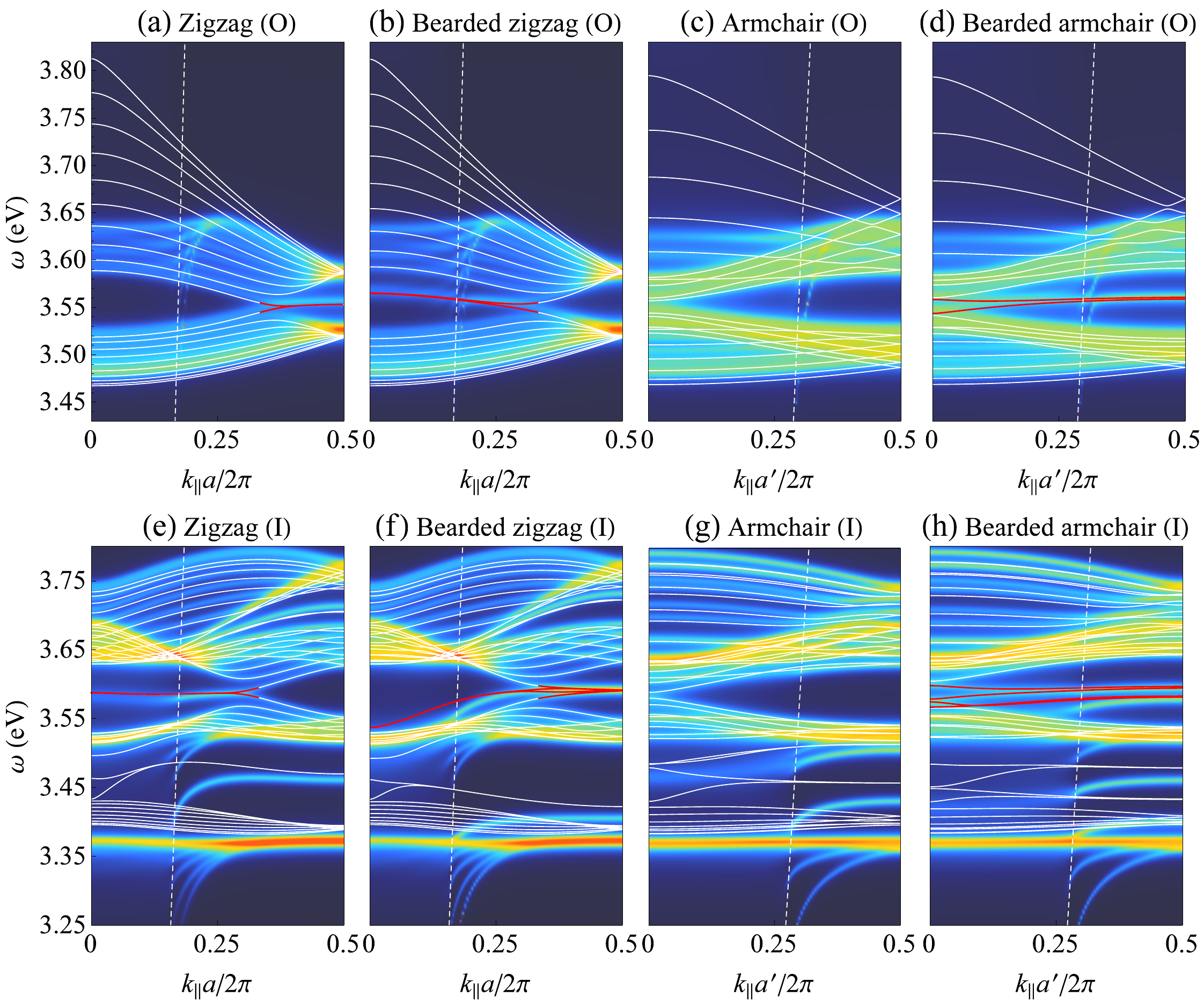}
\caption{\label{fig:4} Out-of-plane (O) and in-plane (I) polarized band structures (white curves) of ribbons including all hoppings under QSA, and the color contours of the extinction cross section where the bright stripes correspond to the band dispersions with retardation and dissipation taken into account. These $8$ subfigures correspond to $4$ different boundaries, i.e. (a)(e) zigzag ribbon, (b)(f) bearded zigzag ribbon, (c)(g) armchair ribbon, (d)(h) bearded armchair ribbon. The white dashed line is the light line in vacuum, the mid-gap edge bands are highlighted in red, and the ribbon width is chosen to be $N = 10$ for all of these eight cases.}
\end{figure}

When long range couplings are taken into consideration, the chiral symmetry is broken, consequently, both bulk and edge bands are distorted, as illustrated by the band structures (white curves) of QSA Hamiltonian including all long range hopping in Fig.~\ref{fig:4}. Nevertheless, the topology of the bands does not change, namely, the projected bulk bands are still degenerate at Dirac points which  are connected by edge bands in corresponding regions. Also, we have broken the inversion symmetry by altering the particle size ratio of two sublattices for these four ribbons, then the gapless surface states split off naturally (see details in Appendix B).

Moreover, since the retardation effect usually influences the band dispersions significantly in plasmonic lattices \cite{weber2004propagation, park2004surface, fung2007plasmonic, zhen2008collective, fung2008analytical}, we have also studies this influence by calculating the band structures with the intact dipole radiation given in Eq.~\ref{eq:1}. The band dispersions with the influence of retardation can be characterized by the loci of the extinction cross section peaks which can be mapped out through the effective polarizability under an external driven field (see Appendix C for detailed derivations) \cite{zhen2008collective}.
On the other hand, the dissipation of nanoparticles also affects the dynamic dispersion and the Q factors of the resonant peaks. Here, we choose the Drude damping term with a small value 0.01 eV to obtain a rather clear band dispersion and also relative high Q factors. As shown in Fig.~\ref{fig:4}, the band structures are in good agreement with QSA results especially far away from the light cone (here only half of the Brillouin Zone is shown due to the symmetry, and the truncation number of hopping is chosen to be 200). The lower out-of-plane bulk band and the two upper in-plane bulk bands are affected little by retardation. However, the upper out-of-plane bulk band and the two lower in-plane bulk bands are dragged down as closing to the light cone and significantly blurred due to coupling with free photons. These observations are also revealed by the multiple scattering based calculations \cite{han2009dirac}. For the mid-gap edge bands corresponding to the flat edge bands in the nearest neighbor quasi-static approximation, outside the light cone, they are also satisfactorily consistent with the QSA results plotted by the red curves in Fig.~\ref{fig:4}. Even inside the light cone, these edge bands are also agree with the results of QSA on the whole, apart from two exceptions. For the in-plane modes in bearded zigzag and bearded armchair ribbons (Fig.~\ref{fig:4}(f) and (h)), a pair of edge bands is dragged down and submerged by the bulk bands in each case, while, in the bearded armchair case, another pair of edge bands is scarcely disturbed by retardation and still overlaps with the QSA bands inside the light cone. It has to be noted that the light cone will move to the right and occupy larger space in the Brillouin zone, when the lattice constant are increased relative to the wavelength. And the occupations of the light cone in the ribbon Brillouin zone are different for ribbons with different boundaries due to their different lengths of periods. As lattice constant increasing, the width of these dipole bands will be compressed, as a results, the narrowed bandwidth compels the resonant peaks to overlap with each other (supposing the damping is unchanged), and eventually the band gaps as well as the edge states are submerged by this compression. Therefore, a much smaller damping is needed to preserve the Q factors of the bands, when increasing the lattice size.

In terms of the above results, we have showed that the essential characters of the HPLs band structures are accurately caught by the simplified nearest neighbor QSA method. According to the simplified QSA results, it is natural to ask why the flat edge states of in-plane polarized modes appear in different ranges than out-of-plane cases, and what determines the existence of these flat edge states. In the previous work \cite{delplace2011zak}, the existence of edge states with a particular $k_\parallel$ in graphene ribbons for arbitrary boundaries has already been related with the Zak phase of $\pi$ bands. This principle of bulk-edge correspondence can be used directly in our system for the out-of-plane cases, but it is invalid to the in-plane cases which have a $4$ dimensional bulk Hamiltonian due to the vector nature of the eigenfields. For this purpose, we will extend this principle into $4$ dimensional vector wave systems with a general form of the Hamiltonian in the next section.

\section{Bulk-edge Correspondence}
\subsection{Relation between number of edge states, Zak phase and Winding number}
For out-of-plane polarized modes, the bulk Hamiltonian possesses a $2$D massless Dirac-like form. According to the bulk-edge correspondence in this kind of systems \cite{delplace2011zak,mong2011edge}, the number of pairs of flat edge states at the mid-gap for a fixed $k_\parallel$, in a ribbon with an arbitrary boundary, can be predicted by the winding number $W(\mathbf{p})$, as well as the Zak phase of either one of the bulk bands divided by $\pi$, $\mathcal{Z}/\pi$. And the Zak phase can be calculated by
\begin{equation}
\mathcal{Z} = i\int_0^{\left| {{\bm{\Gamma} _N}} \right|} {d{k_ \bot }\left\langle {{\psi _i}} \right|} \frac{\partial }{{\partial {k_ \bot }}}\left| {{\psi _i}} \right\rangle  = \pi  \cdot W(\mathbf{p}),
\label{eq:25}
\end{equation}
where the integration is along the path perpendicular to the edge with a fixed $k_\parallel$, $\bm{\Gamma} _N$ is the reciprocal vector perpendicular to the boundary.

In the following, we will derive the existence conditions of the in-plane polarized flat edge states for zigzag boundary as an example, since other boundaries can be analyzed in a similar way. We start from the Bloch states, the eigenstates of Hamiltonian for the infinite system: ${\psi _\mathbf{k}} = {(\psi _\mathbf{k}^A,\psi _\mathbf{k}^B)^T}$. On each site of a unit cell, the dipole moment of this Bloch state is ${(\mathbf{P}_\mathbf{R}^A,\mathbf{P}_\mathbf{R}^B)^T} = {e^{i\mathbf{k} \cdot \mathbf{R}}}{(\psi _\mathbf{k}^A,\psi _\mathbf{k}^B)^T}$ which consists of a common phase factor and the relative amplitudes on two sublattices.

Further, we assume a semi-infinite lattice with a unique edge possesses such edge states that hold the form of Bloch waves but decay exponentially away from the edge to interior. In this case, the wave vector should be extended to complex $\bm{\kappa}  = ({k_\parallel },{\kappa _ \bot })$ with analytical continuation. The parallel component $k_\parallel$ is real and still a good quantum number, while the normal component $\kappa _ \bot$ is complex corresponding to the exponential decay factor. Namely, the phase factor associated with $\kappa _ \bot$ is
\begin{equation}
\xi=\exp\left(i\kappa_\perp\frac{2\pi}{|\bm{\Gamma}_N|}\right)\in\mathbb{C},
\label{eq:26}
\end{equation}
which gives rise to a decay field $\mathbf{P}_\mathbf{n}(\bm{\kappa})\propto \xi^n$ as long as $|\xi|<1$, where $n$ is the number of layers counted from the edge ($n=0$) to interior (see Appendix A for detailed derivation). For zigzag boundary, $\xi=\exp\left(i\kappa_\perp\, 3a_0/2\right)$. Since the edge states $\psi_{\bm{\kappa}}$ maintain the Bloch-like form, they should also obey the eigenequation of bulk hamiltonian $H_{\bm{\kappa}}^{\mathrm{in}}$ with the complex wave vector $\bm{\kappa}$. However, the eigenequation of the edge atoms need to be modified due to the lack of hoppings from sites outside the edge. For zigzag boundary, the eigenequations of the the edge states with Bloch-like form read
\begin{align}
       &{\text{A}\not\in\text{Edge}:} \sum_{i=1}^3\mathbf{\mathord{\buildrel{\lower3pt\hbox{$\scriptscriptstyle\leftrightarrow$}}
\over G}}_i\, e^{i\bm{\kappa}\cdot(\mathbf{t}_3-\mathbf{t}_i)}\psi^B_{\bm{\kappa}}
        = A(\bm{\kappa})\,\psi^B_{\bm{\kappa}}
        =\lambda(\bm{\kappa})\psi^A_{\bm{\kappa}},\label{eq:27}\\
       &{\text{A}\in\text{Edge}:}\, \sum_{i=2,3}\overset\leftrightarrow{\mathbf{G}}_i\,e^{i\bm{\kappa}\cdot(\mathbf{t}_3-\mathbf{t}_i)}\psi^B_{\bm{\kappa}}
      =\lambda(\bm{\kappa})\psi^A_{\bm{\kappa}},\label{eq:28}\\
      &{\text{B}:}\sum_{i=1}^3\mathbf{\mathord{\buildrel{\lower3pt\hbox{$\scriptscriptstyle\leftrightarrow$}}
\over G}}_i\,e^{i\bm{\kappa}\cdot(\mathbf{t}_i-\mathbf{t}_3)}\psi^A_{\bm{\kappa}}
       = A(\bm{\kappa})^\dagger\psi^A_{\bm{\kappa}}=\lambda(\bm{\kappa})\psi^B_{\bm{\kappa}},\label{eq:29}
\end{align}
where $\mathbf{\mathord{\buildrel{\lower3pt\hbox{$\scriptscriptstyle\leftrightarrow$}}
\over G}}_i$ is short for $\mathbf{\mathord{\buildrel{\lower3pt\hbox{$\scriptscriptstyle\leftrightarrow$}}
\over G}}_{xy}(\mathbf{t}_i)$ ($i=1,2,3$), and it is assumed that the edge is terminated at the $A$ sublattice. The missing term in Eq.~(\ref{eq:28}), in comparison with Eq.~(\ref{eq:27}), is corresponds to the hoppings of the missing $B$ atoms cut off by the edge. Here, we are only interested in the flat zero-eigenvalue edge states, i.e., $\lambda(\bm{\kappa})=0$. Substituting $\lambda(\bm{\kappa})=0$ into Eqs.~(\ref{eq:27})--(\ref{eq:29}), then we could obtain the sufficient conditions of the flat edge states
\begin{align}
       &{\text{A}\not\in\text{Edge}:\ } \tilde{A}(k_\parallel,\xi)\,\psi^B_{\bm{\kappa}}
        =0,\label{eq:30}\\
       &{\text{A}\in\text{Edge}:}\ \sum_{i=2,3}\overset\leftrightarrow{\mathbf{G}}_i\,e^{i\bm{\kappa}\cdot(\mathbf{t}_3-\mathbf{t}_i)}\psi^B_{\bm{\kappa}}=0,\label{eq:31}\\
      &{\text{B}:\ } \tilde{A}(k_\parallel,\xi)^\dagger\psi^A_{\bm{\kappa}}=0,\quad \text{subject to }|\xi|<1,\label{eq:32}
\end{align}
where $\tilde{A}(k_\parallel,\xi)\equiv A(\bm{\kappa})$ denotes a complex-valued function of $\xi$ for a special $k_\parallel$. For zigzag ribbon, $\tilde{A}(k_\parallel,\xi)^\dagger=\mathbf{\mathord{\buildrel{\lower3pt\hbox{$\scriptscriptstyle\leftrightarrow$}}
\over G}}_1\,e^{ik_\parallel\frac{a}{2}}\,\xi+\mathbf{\mathord{\buildrel{\lower3pt\hbox{$\scriptscriptstyle\leftrightarrow$}}
\over G}}_2\,e^{ik_{\parallel} a}+\mathbf{\mathord{\buildrel{\lower3pt\hbox{$\scriptscriptstyle\leftrightarrow$}}
\over G}}_3$. To guarantee the self-consistency of Eq.~(\ref{eq:30}) and Eq.~(\ref{eq:31}), $\psi^B_{\bm{\kappa}}$ must vanish: $\psi^B_{\bm{\kappa}}=0$. Furthermore, Eq.~(\ref{eq:32}) with nontrivial solution is equivalent to
\begin{equation}\label{eq:33}
 \det\tilde{A}(k_\parallel,\xi)^\dagger =0,\quad \text{subject to } |\xi|<1.
\end{equation}
Therefore, every zero point of $\det\tilde{A}(k_\parallel,\xi)^\dagger$ inside the unit circle $|\xi|=1$ corresponds to a zero-eigenvalue edge state of the semi-infinite ribbons. For a ribbon with two edges (ribbon width should be large enough so that the coupling between two edges can be neglected), the edge states appear in pairs localized at two sides of the ribbon, i.e. a solution of Eq.~(\ref{eq:33}) corresponds to an edge state on each edge respectively.

As $\xi$ travels along the path $|\xi|=1$, the trajectory of $\det\tilde{A}(k_\parallel,\xi)^\dagger$ is also a closed curve on the complex plane. According to Cauchy's argument principle, the winding number of the trajectory of $\det\tilde{A}(k_\parallel,\xi)^\dagger$ equals to the difference between the number of zeros and poles of $\det\tilde{A}(k_\parallel,\xi)^\dagger$ inside the unit circle. Moreover, $\det\tilde{A}(k_\parallel,\xi)^\dagger$ is analytic and thus possesses no pole inside the closed path. Therefore, with a fixed $k_\parallel$, the winding number of $\det\tilde{A}(k_\parallel,\xi)^\dagger$ is,
\begin{equation}\label{eq:34}
  W\left(\det A^\dagger\right)=\frac{1}{2\pi i }\oint_{|\xi|=1}d\xi\frac{\partial\ln\det\tilde{A}(k_\parallel,\xi)^\dagger}{\partial\xi},
\end{equation}
gives the number of pairs of flat edge states in a finite ribbon system. Since the circle $|\xi|=1$ corresponds to $k_\perp\in[0,\,|\bm{\Gamma}_N|)$ with a fixed $k_\parallel$ which denotes a closed path, marked as $\Gamma_\perp$,  perpendicular to the edge in Brillouin zone (see Appendix A), the winding number of $\det A^\dagger$ also can be integrated with respect to $k_\perp$ alone this path:
\begin{equation}\label{eq:35}
\begin{split}
  W(\det A^\dagger)=&\frac{i}{2\pi }\int_0^{|\bm{\Gamma}_N|}dk_\perp\frac{\partial\ln\det{A}(k_\parallel,k_\perp)}{\partial k_\perp} = \frac{i}{2\pi }\int_0^{|\bm{\Gamma}_N|}dk_\perp \mathrm{Tr}\left(A^{-1}\frac{\partial A}{\partial k_\perp} \right)\\
  =& \text{number of pairs of flat edge states}.
\end{split}
\end{equation}
This relation reveals the correspondence between the existence of the flat edge states and the  bulk Hamiltonian.
Moreover, the existence of the edge states is also related to the Zak phase of each band. Since $A$ is a symmetric matrix, it can be regarded as a symmetric bilinear mapping, or a tensor of $(2,0)$ type. Then, the representation of $A$ in the basis $\{\mathbf{u}_1,\mathbf{u}_2\}$ given by Eq.~(\ref{eq:20}) can be written as
\begin{equation}\label{eq:36}
\begin{split}
 A=\sum_{i,j}\left( \mathbf{u}_i^\dagger A\, \mathbf{u}_j^*
\right)\mathbf{u}_i\otimes\mathbf{u}_j=\sum_{j=1,2}\tilde{\lambda}_j^*\mathbf{u}_j\otimes\mathbf{u}_j=(\mathbf{u}_1,\mathbf{u}_2)\begin{pmatrix}
\tilde\lambda^*_1 & 0 \\
0 & \tilde\lambda_2^*
\end{pmatrix}
\begin{pmatrix}
\mathbf{u}_1\\
\mathbf{u}_2
\end{pmatrix}.
\end{split}
\end{equation}
Thus the bulk Hamiltonian becomes
\begin{equation}\label{eq:37}
 H^{\mathrm{in}}=\sum_{j=1,2}\begin{pmatrix}
 0 &\hspace{-3pt}\tilde{\lambda}_i^*\mathbf{u}_j\otimes\mathbf{u}_j\\
 \tilde{\lambda}_j\mathbf{u}_j^\dagger\otimes\mathbf{u}_j^\dagger & 0
 \end{pmatrix}
 =\sum_{j=1,2}\bm{\Lambda}_j\cdot\bm{\Sigma}_j,
\end{equation}
where $\bm{\Lambda}_j=(\mathrm{Re}\tilde\lambda_j,\mathrm{Im}\tilde\lambda_j,0)^T=\lambda_j(\cos\gamma_j,\sin\gamma_j,0)^T$, and  $\{\bm{\Sigma}_j,\Sigma^0_j\}$ is a set of ``generalized Pauli matrices":
\begin{equation}\label{eq:38}
\begin{split}
&\Sigma^x_j=\begin{pmatrix}0 & \mathbf{u}_j\otimes\mathbf{u}_j\\ \mathbf{u}_j^\dagger\otimes\mathbf{u}_j^\dagger & 0\end{pmatrix},
    \ \Sigma^y_j=i\begin{pmatrix}0 & \hspace{-5pt}-\mathbf{u}_j\otimes\mathbf{u}_j\\ \mathbf{u}_j^\dagger\otimes\mathbf{u}_j^\dagger & 0\end{pmatrix},\\[3pt]
 &\Sigma^z_j=\begin{pmatrix} \mathbf{u}_j\otimes\mathbf{u}_j^\dagger & 0\\ 0 & \hspace{-4pt}-\mathbf{u}_j^\dagger\otimes\mathbf{u}_j \end{pmatrix},\
     \Sigma^0_j=\begin{pmatrix} \mathbf{u}_j\otimes\mathbf{u}_j^\dagger & 0\\ 0 & \mathbf{u}_j^\dagger\otimes\mathbf{u}_j \end{pmatrix},
\end{split}
\end{equation}
which satisfies the Pauli algebra
\begin{equation}\label{eq:39}
 \Sigma^a_j\Sigma^b_j=i\sum_c \epsilon^{abc}\Sigma^c_j +2\delta^{ab}\Sigma^0_j,\quad a,b,c\in\{x,y,z\},
\end{equation}\
and $i\bm{\Sigma}_j/2$ gives a $4$-dimensional representation of $\mathfrak{su}(2)$ Lie algebra. In consideration of Eq.~(\ref{eq:21}), $H^{\mathrm{in}}$ therefore has been separated into two parts which correspond to bands $1$, $4$ and bands $2$, $3$ respectively. By analogy to the out-of-plane case, we can introduce the winding number of $\bm{\Lambda}_j(\mathbf{k})$ (or the complex function $\tilde\lambda_j(\mathbf{k})$), $W(\bm{\Lambda}_j)$, via tracing the trajectory of $\bm{\Lambda}_j$ as $\mathbf{k}$ varying along the closed path with a fixed $k_\parallel$, $k_\perp\in[0,\,|\bm{\Gamma}_N|)$ in the Brillouin zone. On the other hand, the Zak phase of each band reads
\begin{equation}\label{eq:40}
   \mathcal{Z}_j=i\int_0^{|\bm{\Gamma}_N|}d k_\perp \langle\psi_{j,\pm}|\frac{\partial}{\partial k_\perp}|\psi_{j,\pm}\rangle=\frac{\Delta\gamma_j}{2}.
\end{equation}
Obviously, $\mathcal{Z}_j$ equals to $\pi$ times the winding number of $\bm{\Lambda}_j$. In addition, according to Eq.~(\ref{eq:36}), we have
\begin{equation}\label{eq:41}
 \det A=\det U\, \tilde\lambda^*_1\lambda^*_2\,\det U^T=\lambda_1\lambda_2\, e^{-i(\gamma_1+\gamma_2)},
\end{equation}
where $U=(\mathbf{u}_1,\mathbf{u}_2)$ is a unitary matrix satisfying $U=U^\dagger$, and $\det U=-1$. Consequently, the relation between the bulk hamiltonian and the Zak phase (winding number) of each band is:
\begin{equation}\label{eq:42}
 W(\det A)=-\sum_{j=1,2}W(\bm{\Lambda}_j)=-\sum_{j=1,2}\mathcal{Z}_j/\pi.
\end{equation}
Note that $W(\bm{\Lambda}_j)$ and $\mathcal{Z}_j$ can either be positive or negative, and the equality of Eq.~(\ref{eq:42}) is established for arbitrary boundary conditions. By combining Eqs.~(\ref{eq:35}) and ~(\ref{eq:42}), then the main theorem of the bulk-edge correspondence for our system could be obtained.\\
\begin{thm} 
 Consider the in-plane polarized modes of a plasmonic ribbon with honeycomb lattice. For a fixed $k_\parallel$,
 \begin{equation}\label{eq:43}
\begin{split}
    \text{Number of pairs of flat edge states}=W(\det {A}^\dagger)=\sum_{j=1,2}W(\bm{\Lambda}_j)=\sum_{j=1,2}\mathcal{Z}_j/\pi.
\end{split}
\end{equation}
 Namely, the following quantities are equivalent: \\
1. The number of pairs of flat edge states at the mid-gap;\\
2. The winding number of $\det A^\dagger$, $W(\det A^\dagger)$;\\
3. The sum of the winding numbers $W(\bm{\Lambda}_j)$ of bands $1$ and $2$;\\
4. The sum of the Zak phases of bands 1 and 2 divided by $\pi$, $(\mathcal{Z}_1+\mathcal{Z}_2)/\pi$.\\
All of the winding numbers and Zak phases mentioned above are integrated along the path $\Gamma_\perp$ which is perpendicular to the edge and goes through the first Brillouin zone.
\end{thm}
This theorem declares how does the topological properties of the bulk bands determine the existence of the flat edge states and gives their relation quantitatively. Though we only provide the derivation of the theorem for zigzag ribbon in the text, it is not difficult to give a similar proof to other types of ribbons, such as the bearded zigzag ribbon. In the following section, we will study four common boundaries to show the validity of the theorem.

\subsection{Application of the theorem to four common boundaries}
\begin{figure}[!h]
\centering
\includegraphics[width=0.8\linewidth,height=0.712\linewidth]{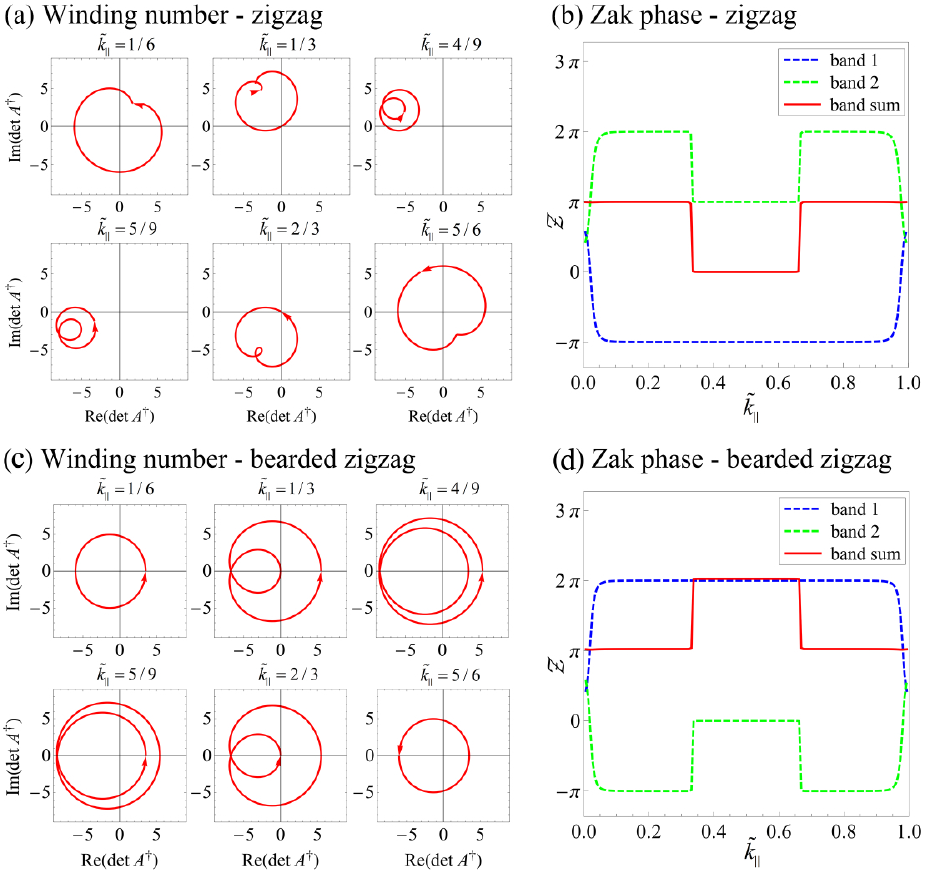}
\caption{\label{fig:5} Winding numbers and Zak phases for zigzag and bearded zigzag ribbons. (a)(c) The trajectories of $\det A^\dagger$ for $6$ discrete $\tilde{k}_\parallel$ as $k_\perp$ varies form $0$ to $|\bm{\Gamma}_N|$, where counterclockwise winding stands for positive direction. (b)(d) Zak phases of bands $1$, $2$ and their summation change with respect to $\tilde{k}_\parallel$.}
\end{figure}

\begin{figure}[!h]
\centering
\includegraphics[width=0.8\linewidth,height=0.712\linewidth]{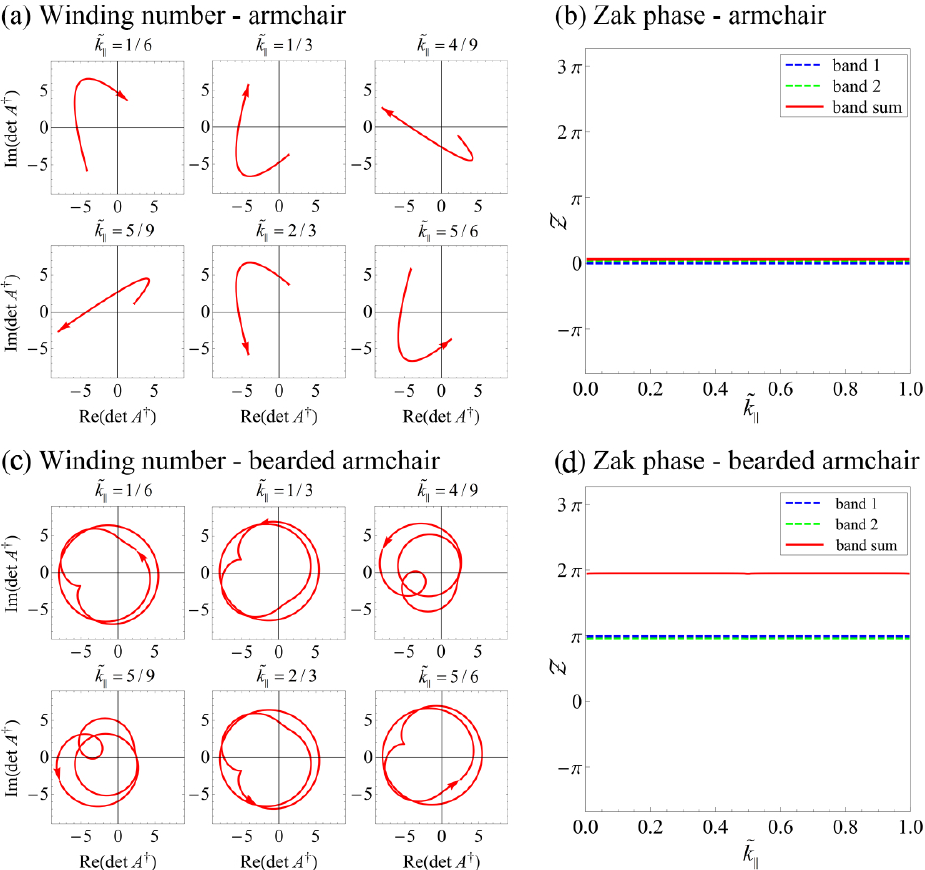}
\caption{\label{fig:6} Winding numbers and Zak phases for armchair and beaded armchair ribbons. (a)(c) The trajectories of $\det A^\dagger$ for 6 discrete $\tilde{k}_\parallel$  as $k_\perp$ varies form $0$ to $|\bm{\Gamma}_N|$. In (a), the trajectories are composed by onward and backward semi-pathes coincident to each other for all $\tilde{k}_\parallel$, thus $W(\det A^\dagger)\equiv 0$. (b)(d) Zak phases of bands $1$, $2$ and their sum in the first Brillouin zone.}
\end{figure}

To verify our conclusion, the Zak phases and winding numbers associated with the in-plane polarized modes are calculated independently for the four typical ribbon boundaries. And we will not show the reduplicative results of out-of-plane polarization as they are similar to graphene ribbon systems \cite{delplace2011zak}. The calculated results of these four ribbons are shown in Fig.~\ref{fig:5}--\ref{fig:6}, the directed loops are the trajectories of $\det A^\dagger(k_\parallel,k_\bot)$ as $k_\bot$ travels alone the path ${\Gamma _ \bot }$ in the Brillouin zone with several fixed $\tilde k_\parallel$ shown on the top of each illustration (${\tilde k_\parallel } ={k_\parallel}a/2\pi $ and ${\tilde k_\parallel } ={k_\parallel}a'/2\pi $ in Fig.~\ref{fig:5} and \ref{fig:6} correspondingly). The number of times that a loop travels counterclockwise around the origin denotes the winding number of $\det A^\dagger$ for the corresponding $\tilde k_\parallel$.

As illustrated in Fig.~\ref{fig:5}(a), two loops of ${\tilde k_\parallel } = 1/3$ and $2/3$ pass through the origin, which correspond to the two phase transition points of the winding number. In Fig.~\ref{fig:5}(b), the sum of the Zak phases of bands $1$, $2$ is non-zero and equals to $\pi$ only in the range of ${\tilde k_\parallel } \in [0,1/3) \cup (2/3,1]$, and it agrees with the winding number of $\det A^\dagger$. This indicates the existence of one pair of flat in-plane polarized edge bands in this range for zigzag ribbon, which is consistent with the band structure illustrated in Fig.~\ref{fig:3}(e). And for the bearded zigzag ribbon, the sum of Zak phases of bands $1$, $2$ equals to $\pi$ in the range of ${\tilde k_\parallel } \in [0,1/3) \cup (2/3,1]$, and equals to $2\pi$ in the range of ${\tilde k_\parallel } \in (1/3,2/3)$, meanwhile, the winding number of $\det A^\dagger$ equals to $1$ and $2$ in these two ranges respectively. This indicates the existence of one pair of flat edge bands in the range of ${\tilde k_\parallel } \in [0,1/3) \cup (2/3,1]$, and two pairs in ${\tilde k_\parallel } \in (1/3,2/3)$ for in-plane polarization.

Fig.~\ref{fig:6}(a) and (b) is the armchair case, it shows the sum of Zak phases of bands $1$, $2$, as well as the winding number of $\det A^\dagger$, equals to $0$ in the whole Brillouin zone. This indicates there are no flat edge states for armchair ribbon. Fig.~\ref{fig:6}(c) and (d) corresponds to bearded armchair ribbon. In this case, the sum of Zak phases of bands $1$, $2$ equals to $2\pi$ in the whole Brillouin zone, meanwhile, the winding number of $\det A^\dagger$ always equals to $2$. This indicates two pairs of flat edge states exist in the whole Brillouin zone.

We have to be noted that the value of Zak phase (winding number) of each band may have a difference of $\pm 2\pi$ from the present result for the different choices of the bulk unit cell, but these differences of bands 1 and 2 always have opposite sign, as a result, the sum of the Zak phases of the two bands is independent of the unit choice. In conclusion, the existence of edge states predicted by winding number or Zak phase are perfectly consistent with the statistics given in Table I for the four common boundary conditions. And these results confirm the validity of Theorem 1.

\section{Distribution of dipole moments}
\begin{figure}[tb]
\centering
\includegraphics[width=1.0\linewidth,height=0.427\linewidth]{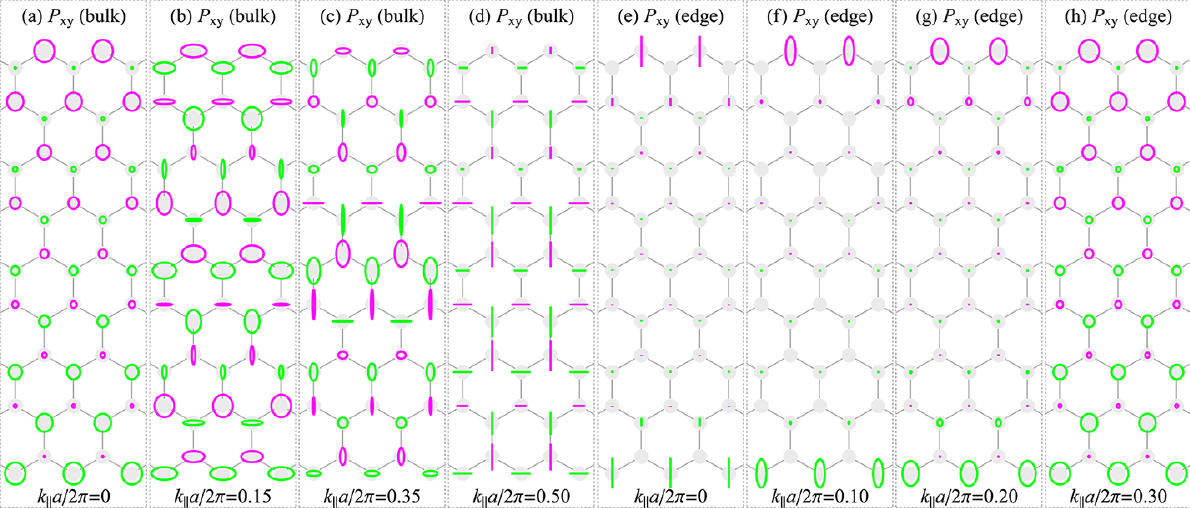}
\caption{\label{fig:7} In-plane dipole moment distributions in the zigzag ribbon with the width $N=9$. The eigenfields are elliptically polarized, and the polarization ellipses are drawn on the corresponding lattice points, where the size of major axis reflects the relative amplitude of the dipole. Green and pink colors are used to distinguish field on two sub-lattices. (a) -- (d) correspond to bulk states at $\omega = 3.676$ eV with changing $k_{\parallel}$, (e) -- (h) correspond to flat edge states at $\omega_0 = 3.568$ eV with changing $k_{\parallel}$.}
\end{figure}

Apart from the band structures, we can also obtain the dipole moment distributions of eigenfields on the lattices. Here the zigzag ribbon of in-plane polarization is shown as an example. As we see in Fig.~\ref{fig:7}, for bulk modes, the dipole moment on each lattice point is comparable for all $k_\parallel$, however, the flat edge modes exponentially decay into the bulk, and only one sublattice, which the outmost atoms belong to, is excited on each side of the ribbon. Besides, the decay rate $|\xi|$ tends to $1$, when $k_\parallel$ approaches Dirac point. In addition, both the bulk modes and the edge modes are elliptically polarized in general, and the eccentricity of polarization ellipse varies in respect to $k_\parallel$. At the boundary of Brillouin zone ($k_\parallel=0$), the edge mode is linearly polarized with the direction perpendicular to the edge (Fig.~\ref{fig:7}(e)), whereas the edge mode tends to circular polarization when $k_\parallel$ goes toward the projected Dirac point (Fig.~\ref{fig:7}(h)). And this effect reflects the vector nature of in-plane edge modes, and thus is a distinctive property different from out-of-plane cases. Hitherto, with our method, both the characters of dipole moment distribution and topological properties of the in-plane polarized mid-gap edge modes are clearly depicted.

\section{Conclusion}
In this paper, we have investigated the band structures and edge states in honeycomb plasmonic lattices with different boundaries. We have proposed a definite correspondence between the Zak phases of bulk bands, the winding number associated with the bulk Hamiltonian, and the existence of flat edge states with in-plane polarization for arbitrary ribbon boundaries, and further verified it through four typical ribbons. We emphasis that the vector nature of the in-plane polarized modes in our system makes the existence conditions of edge states differ from the out-of-plane polarized modes, the $\pi$ electronic excitations in graphene, and all of the systems possessing a $2$ dimensional Dirac-like Hamiltonian. Therefore, the theorem of bulk-edge correspondence obtained in this paper is a novel extension to previous works.

Besides, we also give some necessary remarks to our results. Firstly, we have adopted several assumptions and approximations, i.e. we suppose a proper scale of the system parameters so that the CDM and QSA are applicable, and we also omit the dissipation in the system. Nevertheless, according to our comparison between the QSA and the retardation-affected results, the simplified QSA model catches the main characters of the real dipole bands accurately. Secondly, this bulk-edge correspondence is derived in tight-binding approximation with only nearest neighbor interaction. When long range interactions are taken into consideration, both the bulk bands and edge bands will be distorted. In spite of some changes in band structures, the topological properties do not change (see Appendix B). Thirdly, we only give a proof of the theorem for zigzag in the text. Although similar proofs are easily obtained case-by-case for some other boundary conditions, a general proof for arbitrary boundaries is still hard, therefore, Eq.~(\ref{eq:43}) is still a hypothesis to some extent. Fourthly, our theorem is only applicable to predict the flat edge bands at the mid-gap, but is helpless to predict the existence of the bent edge bands between bands $1$, $2$ and bands $2$, $3$ of in-plane polarized modes in these ribbons. It is still an open question whether there is a topological invariant corresponding to these bent edge states.

Finally, the model proposed here is not limited to photonic systems, and we believe it can also be applied to other 2D wave systems which have an arbitrary even order Hamiltonian with chiral symmetry. On the other hand, the honeycomb plasmonic lattice not only is a beautiful analogy of classical wave to graphene but also owns additional interesting properties which may be easier to be realized experimentally. We hope our work could inspire further theoretical and experimental researches of similar systems.

\section*{Acknowledgements}
We thank Prof. Zhao-Qing Zhang, Dr. Min Yang and Dr. Shubo Wang for useful discussions. This work was supported by a grant from the Research Grants Council of the Hong Kong (Project No. AoE/P-02/12) and National Natural Science Foundation of China (NSFC) (11304038).

\section*{Appendix A: Basis vector for different boundaries}
In this Appendix, we briefly review how to select a appropriate pair of bases of the honeycomb Bravais lattice for different ribbon boundaries \cite{delplace2011zak}. We choose a pair of vectors $(\mathbf{T},\mathbf{N})$ to be the set of bases corresponding to the ribbon with the periodic vector $\mathbf{T}$, where $\mathbf{T}=m\bm{a}_1+n\bm{a}_2$ and $\mathbf{N}=m'\bm{a}_1+n'\bm{a}_2$. The bases should satisfy ${\bm{a}_1} \times {\bm{a}_2} = \mathbf{T} \times \mathbf{N}$ which leads to $ mn' - nm' = 1$, and $(n',m')$ can be thus determined (the choice of $(n',m')$ is not unique). For zigzag ribbon, $\mathbf{T}_z,\mathbf{N}_z$ are just the original bases $(\bm{a}_1,\bm{a}_2)$. The reciprocal vector bases $({\bm{\Gamma _T}},{\bm{\Gamma _N}})$ corresponding to $(\mathbf{T},\mathbf{N})$ satisfy
\begin{equation}
\left\{ \begin{array}{l}
{\bm{\Gamma _T}} \cdot \mathbf{T} = 2\pi \\
{\bm{\Gamma _N}} \cdot \mathbf{T} = 0
\end{array} \right.,\left\{ \begin{array}{l}
{\bm{\Gamma _T}} \cdot \mathbf{N} = 0\\
{\bm{\Gamma _N}} \cdot \mathbf{N} = 2\pi
\end{array}. \right.
\label{eq:A1}
\end{equation}
\begin{figure}[tb]
\centering
\includegraphics[width=0.85\linewidth,height=0.284\linewidth]{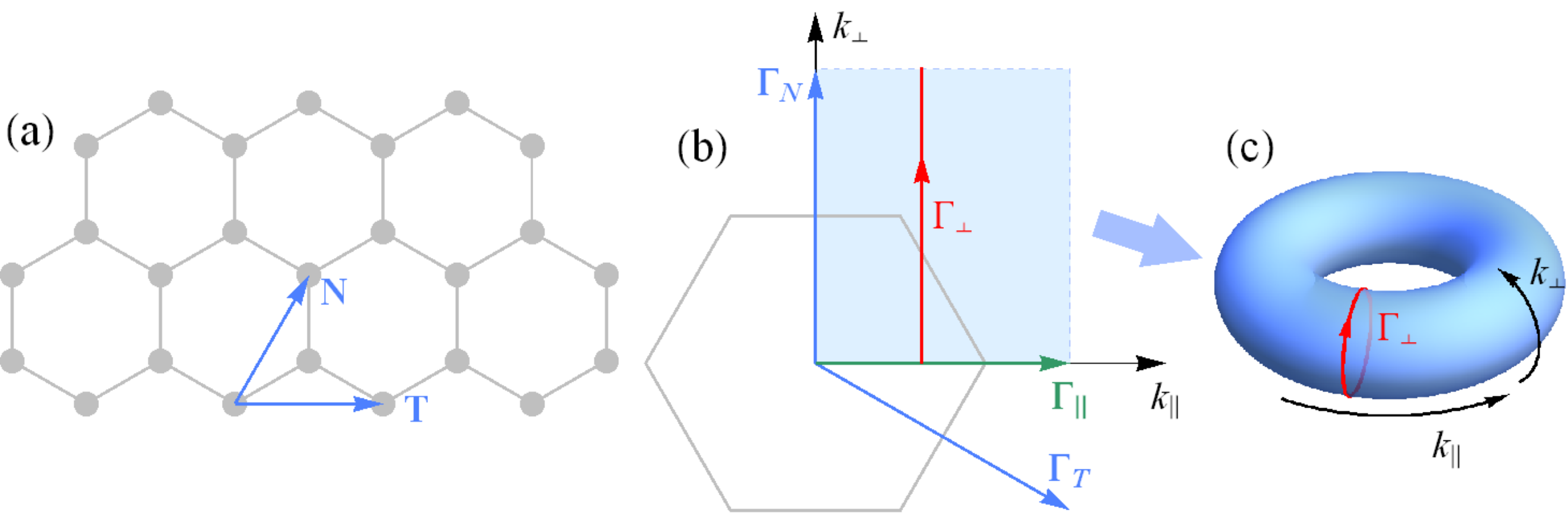}
\caption{\label{fig:app1} Schematic diagram of the alternative bulk Brillouin zone corresponding to zigzag ribbon. (a) The bases $(\mathbf{T},\mathbf{N})$ of Bravais lattice correspond to zigzag ribbon. (b) The bases $(\bm{\Gamma}_T,\bm{\Gamma}_N)$ of reciprocal lattice agree with $(\mathbf{T},\mathbf{N})$. The light blue area is the alternative bulk Brillouin zone constructed by $(\bm{\Gamma}_T,\bm{\Gamma}_\parallel)$, where $\bm{\Gamma}_\parallel$ forms the 1 dimensional Brillouin zone of zigzag ribbon. The path $\Gamma_\perp$ is perpendicular to the edge with a fixed $k_\parallel$. (c) The periodic boundary condition makes the alternative Brillouin zone into a torus. Then $\Gamma_\perp$ maps to a loop on the torus.}
\end{figure}
Then, we have
\begin{equation}
\left\{ \begin{array}{*{20}{l}}
{\bm{\Gamma _T}} = n'{{\bm{b}}_1} - m'{{\bm{b}}_2}\\
{\bm{\Gamma _N}} =  - n{{\bm{b}}_1} + m{{\bm{b}}_2}
\end{array}, \right.
\label{eq:A2}
\end{equation}
where $(\bm{b}_1, \bm{b}_2)$ are the reciprocal bases corresponding to $(\bm{a}_1,\bm{a}_2)$. $\bm{\Gamma}_T$ and $\bm{\Gamma}_N$ form a parallelogram that can be regarded as an alternative selection of Brillouin zone in the reciprocal space. Nevertheless, $\bm{\Gamma}_N$ and an arbitrary vector $\bm{\Gamma}$ can also forms a bulk Brillouin zone as long as $\bm{\Gamma}\times\bm{\Gamma}_N = \bm{\Gamma}_T\times\bm{\Gamma}_N$. We demand $\bm{\Gamma}$ is parallel to the edge, and obtain $\bm{\Gamma}_\parallel=(\bm{\Gamma}_T\cdot\hat{\mathbf{t}})\hat{\mathbf{t}}$. And $|\bm{\Gamma}_\parallel|=2\pi/|\mathbf{T}|$ is precisely the length of the Brillouin zone of the ribbon. The relations of these bases for zigzag ribbon is shown in Fig.~\ref{fig:app1}. For an arbitrary wave vector $\mathbf{k}$, it can be decomposed with respect to the directions of the two reciprocal bases, $\mathbf{k}=\mathbf{k}_{\Gamma_T}+\mathbf{k}_{\Gamma_N}$, or equivalently, decomposed into the two components parallel and perpendicular to the edge of the ribbon, $\mathbf{k}=k_\parallel \hat{\mathbf{t}} + k_\perp \hat{\mathbf{n}}$, where $\hat{\mathbf{t}}$ and $\hat{\mathbf{n}}$ are the unit vectors parallel and perpendicular to the edge. The relations between the two kinds of components are
\begin{equation}
\left\{ {\begin{array}{*{20}{l}}
{k_\parallel } = {{\mathbf{ k }}_{{\Gamma _T}}} \cdot \hat{\mathbf{t}} \\
{k_ \bot } = \left|{\mathbf{k}}_{{\Gamma _N}} \right| + {{\mathbf{k}}_{{\Gamma _T}}} \cdot \hat{\mathbf{n}}
\end{array}.} \right.
\label{eq:A3}
\end{equation}
In all eigenequations above, the wave vector appears only in the fixed term $\mathbf{k}\cdot\mathbf{R}$ of the phase, where $\mathbf{R}=l_1\mathbf{T}+l_2\mathbf{N}$ is an arbitrary lattice vector. This term can be decomposed as follows
\begin{equation}
\mathbf{k} \cdot \mathbf{R} = {k_\parallel }[l_1\left| \mathbf{T} \right| + l_2(\mathbf{N} \cdot \hat{\mathbf{t}})] + l_2{k_ \bot }\frac{{2\pi }}{{\left| {{\bm{\Gamma _N}}} \right|}}.
\label{eq:A4}
\end{equation}
So in Wannier picture, the component of Bloch eigenfunction on lattice point $\mathbf{R}$  can be written as
\begin{equation}
\mathbf{P}_{\mathbf{R}} =e^{i\mathbf{k}\cdot\mathbf{R}}\psi_{\mathbf{k}}= {({e^{i{k_{\perp}}\frac{{2\pi }}{{\left| {{\bm{\Gamma _N}}} \right|}}}})^{l_2}}{e^{i{k_\parallel }[l_1\left| \mathbf{T} \right| + l_2(\mathbf{N} \cdot \hat{ \mathbf{t}})]}} \psi_{\mathbf{k}}.
\label{eq:A5}
\end{equation}
For the edge states in ribbon systems, $k_\perp$ should be replaced by the complex number $\kappa_\perp$, and the first phase term in Eq.~(\ref{eq:A5}) corresponds to the decay of the fields from the edge to the interior, $l_2$ is the layer number at $\mathbf{R}$. Thus, the decay rate of the edge state is $\xi=\exp(i\kappa_\perp\frac{2\pi}{|\bm{\Gamma}_N|})$. For zigzag ribbon, $\bm{\Gamma}_N=\mathbf{b}_2$, so we have $|\bm{\Gamma}_N|=4\pi/(3a_0)$, and the decay rate $\xi=\exp(i\kappa_\perp3a_0/2)$. The unit circle $|\xi|=1$ on the complex plane corresponds to the path $k_\perp \in [0,|\bm{\Gamma}_N|)$ with a fixed $k_\parallel$. Since the periodicity of the lattice makes the Brillouin zone composed by $(\bm{\Gamma}_\parallel,\bm{\Gamma}_N)$ be a torus, the path, $k_\perp \in [0,|\bm{\Gamma}_N|)$ with a fixed $k_\parallel$, forms a closed path, marked as $\Gamma_\perp$, on the torus, as shown in Fig.~\ref{fig:app1}(c).

\section*{Appendix B: Long range interaction and symmetry breaking effect}
The chiral symmetry will be destroyed if we add second neighbour hopping terms (or more long range hopping terms) in the tight-binding Hamiltonian. Here we consider all long range hoppings to exhibit how the long-range interactions affect the band structures in our system. When long range hopping terms are included, the diagonal terms of bulk Hamiltonian becomes non-zero
\begin{equation}
H_{\mathbf{k}} = \left( {\begin{array}{*{20}{c}}
{{H_{AA}}}&{{H_{AB}}}\\
{{H_{BA}}}&{{H_{BB}}}
\end{array}} \right),
\label{eq:B1}
\end{equation}
where $H_{AA}$ and $H_{BB}$ are contributed by the hopping terms of $A$ to $A$ and $B$ to $B$ respectively, while $H_{AB}$ and $H_{BA}$ are composed of hopping terms of $B$ to $A$ and $A$ to $B$ respectively. Since the QSA Green tensor decays with distance in proportion to $r^{-3}$, the band dispersions converge rapidly with respect to the coupling distance.  Here, we add the hopping terms from 400 neighbor ribbon unit cells (left and right 200 respectively) to a certain unit cell into the ribbon Hamiltonian, and calculate the band structures including these long-range interactions to approach the real system.
\begin{figure}[tb]
\centering
\includegraphics[width=0.8\linewidth,height=0.68456\linewidth]{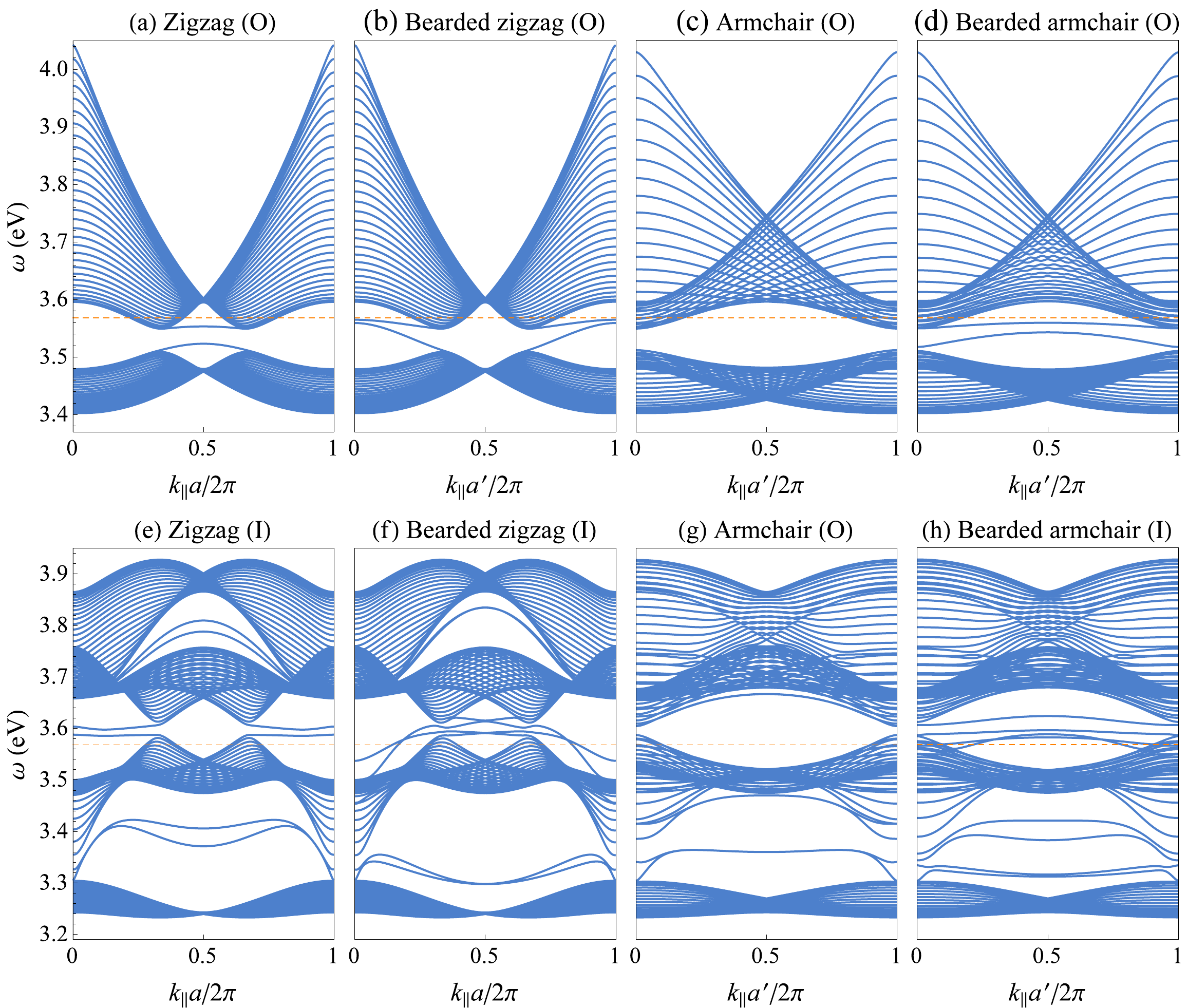}
\caption{Out-of-plane (O) and in-plane (I) polarized QSA band structures with all long range interactions in these four ribbons, i.e. (a)(e) zigzag ribbon, (b)(f) bearded zigzag ribbon, (c)(g) armchair ribbon, (d)(h) bearded armchair ribbon with the defined ratio $s=3$.}
\label{fig:app2}
\end{figure}
As shown in Fig.~\ref{fig:4}, the band structures are distorted by the long-rang interactions. For out-of-plane modes, the upper bulk band is stretched, while the lower bulk band is compressed as comparing to the situation with solo nearest neighbor interactions. Besides, the mid-gap edge bands are not flat any more and the corresponding frequency shifts a little from $\omega_0$. For the in-plane polarization, the band structure also loses its mirror symmetry with respect to the mid-gap plane. Nevertheless, the topological properties of the bands do not change, concretely, the upper and lower bulk bands still degenerate at the Dirac points, despite their small shifts, and the mid-gap edge bands still appear in the ranges predicted by our theorem and connect the projected Dirac points, so the present result further illustrates our method works.

As the ribbon is made up of two sublattices, we can break the inversion symmetry by changing the size ratio of $A$ type particles to $B$ type particles. Here we define a ratio factor
$s = r_{A}^3 / r_{B}^3$, where $r_A$ and $r_B$ are the radii of $A$ and $B$ types of particles respectively. Apparently, $s$ can be assigned arbitrary value, but we have to care about that the ratio must be reasonable so that the coupled dipole approximations are appropriate. In this sense, the Hamiltonian becomes
\begin{equation}
H_\mathbf{k} = \left( {\begin{array}{*{20}{c}}
{{H_{AA}}}&{{H_{AB}}}\\
{{s\cdot H_{BA}}}&{{s\cdot H_{BB}}}
\end{array}} \right).
\label{eq:B2}
\end{equation}

Then we can map out the band structures in the similar way as before, the result is shown in Fig.~\ref{fig:app2}. It is known that if either inversion symmetry or time-reversal symmetry is broken, the degeneracy at the Dirac points will immediately be lifted in 2D band structures according to the Von Neumann$\textrm{-}$Wigner theorem \cite{raghu2008analogs, haldane2008possible}. In contrast to graphene, the degeneracy will not open if we only consider nearest interactions, unless the second hopping influence is also included into the Hamiltonian, because there are no on-site energy terms in our eigen equation. For ribbon band structures, the original degenerate pairs of edge bands will completely separate into two gaped bands contacting the upper and lower bulk bands respectively, however, their topology is trivial since the Chern numbers of these bulk bands are zero when merely breaking the inversion symmetry.

\section*{Appendix C: Dynamic resonance with retardation}
The method used to calculate the band structures including the effect of retardation is mainly based on Ref.~\cite{zhen2008collective}. Here, we give a brief self-contained introduction. We still apply the dipole approximation for particles but treat the particle interactions between dipoles with the complete dipole field given in Eq.~(\ref{eq:1}) which can be expressed as $\mathbf{E}(\mathbf{r},\omega)=\frac{1}{4\pi\varepsilon_0}\matr{G}(\mathbf{r},\omega)\cdot\mathbf{P}(\omega)$, where $\matr{G}(\mathbf{r},\omega)$ is used to denote the Green tensor of complete dipole field in this appendix but not the QSA Green tensor used in the main text. Then, the coupled dipole equation driven by an external field $\mathbf{E}^{\mathrm{driv}}(\mathbf{r})$ reads
\begin{equation}\label{eq:C1}
   \sum_{\mathbf{R}'}\matr{M}_{\mathbf{R},\mathbf{R}'}\cdot\mathbf{P}_{\mathbf{R}'}=\mathbf{E}^{\mathrm{driv}}_{\mathbf{R}},
\end{equation}
where
\begin{equation}\label{eq:C2}
  \matr{M}_{\mathbf{R},\mathbf{R}'}(\omega)=
  \left(\frac{1}{\alpha(\omega)}\matr{I}_{6\times6}+\matr{I}_{2\times2}\otimes\matr{G}(\Delta\mathbf{R},\omega)\right)\delta_{\mathbf{R},\mathbf{R}'}-
  \begin{pmatrix}
    \matr{G}(\Delta\mathbf{R},\omega) & \matr{G}(\Delta\mathbf{R}-\mathbf{t}_3,\omega)\\
     \matr{G}(\Delta\mathbf{R}+\mathbf{t}_3,\omega)& \matr{G}(\Delta\mathbf{R},\omega)
  \end{pmatrix},
\end{equation}
with $\Delta\mathbf{R}=\mathbf{R}-\mathbf{R}'$, and the following substitution should be employed for radiation correction
\begin{equation}\label{eq:C3}
   \frac{1}{\alpha(\omega)}\rightarrow \frac{1}{\alpha(\omega)}-i\frac{2\omega^3}{3c^3}.
\end{equation}

Accordingly, the coupled dipole equation for ribbon systems can be written as
\begin{equation}\label{eq:C4}
  \sum_{l'}\tilde{\mathbf{M}}_{l,l'}\cdot\tilde{\mathbf{P}}_{l'}=\tilde{\mathbf{E}}^{\mathrm{driv}}_{l}.
\end{equation}
Here, $l,l'$ denote the indices of ribbon unit cells, $\tilde{\mathbf{P}}_l=(\mathbf{P}_{\mathbf{R}(l,1)},\cdots,\mathbf{P}_{\mathbf{R}(l,N)})^T$, $\tilde{\mathbf{E}}^{\mathrm{driv}}_l=(\mathbf{E}^{\mathrm{driv}}_{\mathbf{R}(l,1)},\cdots,\mathbf{E}^{\mathrm{driv}}_{\mathbf{R}(l,N)})^T$  ($N$ is the layer number of the ribbon) ,
 and every $\tilde{\mathbf{M}}_{l,l'}=\tilde{\mathbf{M}}(l-l')$ is a $6N\times 6N$ matrix satisfying $\big(\tilde{\mathbf{M}}_{l,l'}(\omega)\big)_{m,m'}=\matr{M}_{\mathbf{R}(l,m),\mathbf{R}(l',m')}$, where $m,m'$ are the layer indices, and a pair of numbers $(l,m)$ locates a certain bulk unit cell $\mathbf{R}(l,m)$. In Bloch representation, the coupled dipole equation turns into $\tilde{\mathbf{M}}(k,\omega)\cdot\tilde{\mathbf{P}}_k=\tilde{\mathbf{E}}^{\mathrm{driv}}_k$
 with $\tilde{\mathbf{M}}(k,\omega)=\sum_{\Delta l}\tilde{\mathbf{M}}(\Delta l,\omega)e^{-ik\Delta l}$. This equation can be alternatively expressed as
 \begin{equation}\label{eq:C5}
  \tilde{\mathbf{P}}_k=\tilde{\mathbf{M}}(k,\omega)^{-1}\cdot\tilde{\mathbf{E}}^{\mathrm{driv}}_k=\matr{\alpha}_{\mathrm{eff}}(k,\omega)\cdot\tilde{\mathbf{E}}^{\mathrm{driv}}_k.
 \end{equation}
 As a result, $\tilde{\mathbf{M}}(k,\omega)^{-1}$ serves as the effective polarizability $\matr{\alpha}_{\mathrm{eff}}(k,\omega)$ of the ribbon lattice in response to the driven field. And each eigenvalue ${\alpha}^i_{\mathrm{eff}}(k,\omega)$ of $\matr{\alpha}_{\mathrm{eff}}(k,\omega)$ indicates one forced oscillation mode of the lattice at the driven frequency $\omega$ and Bloch vector $k$. In terms of optical theorem, the total extinction cross section of all modes takes the following form
 \begin{equation}\label{eq:C6}
   C_{\mathrm{ext}}(k,\omega)=\frac{4\pi\omega}{c}\sum_i {\alpha}^i_{\mathrm{eff}}(k,\omega)=\frac{4\pi\omega}{c}\ \mathrm{Tr}\left( \matr{\alpha}_{\mathrm{eff}}(k,\omega)\right).
 \end{equation}
 Since the loci of the total extinction cross section peaks just correspond to the collective resonances of the plasmonic lattice, the band dispersion including the whole dipole interactions and retardation can accordingly be exhibited by the loci of the peaks of $C_{\mathrm{ext}}$. Besides, $\tilde{\mathbf{M}}(k,\omega)$ is also composed by the direct sum of the in-plane polarization subspace and the out-of-plane polarization subspace, therefore, the band structures of in-plane modes and out-of-plane modes still can be investigated independently by this method.

\section*{References}
\providecommand{\newblock}{}

\end{document}